\PassOptionsToPackage{numbers,sort&compress}{natbib}

\documentclass[aps, preprintnumbers,amsmath,amssymb,prd, notitlepage,nofootinbib, superscriptaddress]{revtex4-2}

\usepackage{amsfonts,amssymb,amsmath}
\usepackage{textcomp}
\usepackage{gensymb}
\usepackage{color}
\usepackage{graphicx}
\usepackage{multirow}
\usepackage[normalem]{ulem}
\usepackage[utf8]{inputenc}
\usepackage{orcidlink}

\usepackage{aas_macros}

\usepackage{hyperref}
\usepackage{array}

\usepackage{booktabs}

\usepackage[normalem]{ulem}

\definecolor{lightgray}{gray}{0.95}

\usepackage{tikz}

\usetikzlibrary{shapes.geometric, arrows, positioning, decorations.pathreplacing, calc}

\newcommand{\Planck}{{\it Planck}~}
\newcommand{\be}{\begin{equation}}        
\newcommand{\ee}{\end{equation}}

\usepackage{soul}
\definecolor{pastelyellow}{HTML}{FFF9A6}
\sethlcolor{pastelyellow}

\usepackage[normalem]{ulem}
\usepackage{xcolor}

\definecolor{milesgreen}{HTML}{008060}
\newcounter{milescomment}

\begin{document}

\title{Single Frequency CMB Foreground Removal with Inter-scale Machine Learning}

\author{Helen Shao}
\orcidlink{0000-0002-0152-6747}
\affiliation{Department of Astronomy, Harvard University}
\affiliation{Institute of Astronomy, University of Cambridge}
\email{helen.shao@cfa.harvard.edu}

\author{Fiona McCarthy}
\affiliation{Department of Applied Mathematics and Theoretical Physics, University of Cambridge}
\affiliation{Kavli Institute for Cosmology, University of Cambridge}
\affiliation{Center for Computational Astrophysics, Flatiron Institute, New York, NY, USA 10010}

\author{Blake D.~Sherwin}
\affiliation{Department of Applied Mathematics and Theoretical Physics, University of Cambridge}
\affiliation{Kavli Institute for Cosmology, University of Cambridge}

\author{Miles Cranmer}
\affiliation{Institute of Astronomy, University of Cambridge}
\affiliation{Department of Applied Mathematics and Theoretical Physics, University of Cambridge}
\affiliation{Kavli Institute for Cosmology, University of Cambridge}

\author{Carlos Herv\'ias-Caimapo}
\orcidlink{0000-0002-4765-3426}
\affiliation{Instituto de Astrof\'isica and Centro de Astro-Ingenier\'ia, Facultad de F\'isica, Pontificia Universidad Cat\'olica de Chile, Av. Vicu\~na Mackenna 4860, 7820436 Macul, Santiago, Chile}

\date{\today}

\begin{abstract}
Accurate measurements of Cosmic Microwave Background (CMB) B-mode polarization, a key probe of inflationary physics, are hindered by complex Galactic dust foregrounds. Traditional foreground removal with Internal Linear Combination (ILC) fully preserves the primordial signal but requires multi-frequency data and is limited to  two-point statistics. We present a novel way to estimate and remove foregrounds at a \textit{single} frequency using signal-preserving machine learning that leverages inter-scale correlations in the foregrounds. Using \textsc{DustFilaments} simulations of Galactic dust maps, we train CNNs to reconstruct large-scale foregrounds ($\ell < 200$) from small-scales ($\ell > 200$). We quantify the effectiveness of foreground removal with the residual foreground power, $f_{\rm{resid}}$, which gives the fraction of foreground power remaining after removal. Predictions using only small-scale $B$-modes achieve $f_{\rm{resid}}\simeq 0.704$, while including temperature and $E$-modes decreases it to $f_{\rm{resid}} \simeq 0.376$. These results are still higher than the spatial ILC's $f_{\rm{resid}} \simeq 2.69\times10^{-3}$, which leverages multi-frequency data at Simons-Observatory-like frequencies. However, a hybrid network that leverages \textit{both} multi-frequency and inter-scale correlations achieves much better cleaning, with $f_{\rm{resid}}\simeq 4.71\times10^{-4}$ when using $B$-mode inputs alone, and $3.62\times10^{-4}$ when using temperature and $E/B$-mode inputs. This network achieves a residual power of $\sim 7\times$ lower than that of spatial ILC, while inheriting ILC's signal-preserving property. Notably, it is also $\sim 2$--$3\times$ lower than that of a network that only uses multi-frequency inputs, demonstrating that correlations across scale are not redundant with correlations across frequency. Hence, inter-scale and cross-component correlations can jointly provide useful foreground information beyond frequency dependence, making our techniques complementary to multi-frequency foreground removal. However, we emphasize that these results are achieved only for the \textsc{DustFilments} model and find that network generalization to other simulations remains a key challenge for robust ML-based foreground removal. 
\end{abstract}

\maketitle

\section{Introduction}\label{sec:introduction}
Diffuse Galactic foregrounds are a major impediment to extracting early-Universe information from polarized CMB observations~\citep{planck_dust, keck2025_comp_sep}. Dominant across all observed bandwidths, these foregrounds range from synchrotron~\citep{smoot_synch} and free-free emission~\citep{Dickinson_2003} at low frequencies ($\nu \sim 10-100 $ GHz) to thermal dust radiation~\citep{planck15_diffuse, bennett_wmap} at high frequencies ($\nu \gtrsim 100 $ GHz). This is particularly challenging for searches for primordial gravitational waves, a direct probe of inflation~\citep{Seljak_1997, kamionkowski1997detecting}, which imprint a distinctive signature in CMB B-mode polarization on large angular scales (angular multipole $\ell < 200$)~\citep{seljak1997signature}. The amplitude of this primordial B-mode signal is parametrized by the tensor-to-scalar ratio, $r$, which quantifies the energy scale of single-field inflation. Current and upcoming experiments aim to constrain $r$ with increasing precision, with near-future targets of order $\sigma(r)\sim10^{-3}$~\citep{Tristram2022}. However, this faint potential imprint is obscured by the gravitational lensing of E-mode polarization into B-mode polarization~\citep[e.g.,][]{LEWIS_2006}, as well as the B-modes of polarized Galactic foregrounds. Notably, the amplitudes of the latter component are over two orders of magnitude larger than those of any potential primary signal on all observed scales~\citep{planck2016foregrounds, planck2018componentsep}. Foreground modeling and removal are therefore critical to improving the statistical significance of a primordial B-mode detection and maximizing the scientific return of existing and future CMB experiments, including BICEP/KECK~\citep{Ade_2022}, the Simons Observatory (SO)~\citep{SimonsObservatory2019,2025arXiv251215833T}, and LiteBIRD~\citep{LiteBIRD2022}.

{One of the most widely used methods for this objective is the Internal Linear Combination (ILC), which exploits cross-frequency correlations in the spatial fluctuations of the various galactic foreground components, and their different spectral energy distributions (SEDs) compared to the CMB~\citep{1992ApJ...396L...7B,bennett2003firstyear, eriksen2004ilc, Delabrouille_2008}.} The ILC and its variants estimate the primary CMB as a weighted linear combination of multi-frequency maps, producing a minimum-variance solution that is optimal for Gaussian, isotropic foregrounds~\citep{eriksen2004ilc}.\footnote{Some anisotropy can be incorporated by performing a spatially-varying ILC over the sky.} Importantly, since it is constrained to preserve the primary CMB signal, the ILC solution is \textit{signal-preserving}. This means that while the ILC reconstruction may contain residual foregrounds, it fully recovers the primordial cosmological information independent of uncertainties and incomplete knowledge of the complex astrophysical and Galactic sources, thus making it reliable for analysis and parameter inference. Due to this analytic tractability and signal-preserving property, the ILC has been widely used across generations of CMB experiments, including \textit{COBE}~\citep{cobe,1992ApJ...396L...7B}, \textit{WMAP}~\citep{bennett_wmap}, \textit{Planck}~\citep{planck15_diffuse}, and the Atacama Cosmology Telescope (ACT)~\citep{ACTPol,2020PhRvD.102b3534M,2024PhRvD.109f3530C, coulton2023atacamacosmologytelescopehighresolution}, and is one of the planned B-mode analysis pipelines for SO~\cite{2024A&A...686A..16W,2026arXiv260414088C}.\footnote{Other CMB collaborations, such as the South Pole Telescope (SPT), sometimes use similar optimal linear combinations where a covariance matrix is not estimated directly from the data (as it is for ILC) but instead from simulations of the sky~\cite{2026PhRvD.114b3534M,2022ApJS..258...36B}.} 

However, the ILC has some critical drawbacks. First, it relies only on inter-frequency correlations of the foregrounds, thereby requiring multi-frequency data to resolve the SEDs of the various foreground components. Second, the ILC solution only optimizes for second-order statistics, neglecting higher-order correlations that are abundant in Galactic foregrounds~\citep{planck2016interstellar, Vansyngel_2018}. Specifically, anisotropies are sourced by the local dust structures and filaments in the Milky Way that align with the Galactic magnetic field, leading to polarization that is coupled across scale and with emission strength increasing nearer the Galactic plane. The turbulent processes governing the trajectories of Galactic dust also introduce non-linear physics that correlate these structures in non-Gaussian manners~\citep{planck2016foregrounds, Kritsuk_2018, Kandel_2017, 2017ApJ...839...91C, Abril_Cabezas_2023}. Hence, the ILC solution may leave residual foregrounds that are not well captured by the covariance matrix. 

Due to these limitations, recent works have turned to machine learning (ML) {or information from higher-order statistics to perform CMB component separation~\citep{petroff2020, likhit2025generativereconstructionlowellcmb, Puglisi_2022, mccarthy24_ml, Bonjean_2024, casas2025recoveringcmbpolarizationmaps, yan2024cmbfscnncosmicmicrowavebackground,Yan_2023,2024arXiv241101233Y,2025JCAP...08..058A,2025ApJS..281...47Y,2025arXiv250900139M,2025arXiv250320774S, bachar2026robustnessneuralnetworkscmb, Jeffrey_2021,R_galdo_Saint_Blancard_2023, Aylor_2020, duivenvoorden2025}}. Typically, ML techniques are used for this task in a supervised manner (see, however,~\citep[]{Bonjean_2024} for a self-supervised example) where neural networks are trained on foreground simulations to learn their non-linear physics and possibly capture the non-Gaussian features missed by ILC. However, a critical risk is that since neural networks are highly sensitive to the inherent assumptions, uncertainties and approximations of the training data (e.g., simulations of specific foreground models), mismatches between simulations and real data can introduce bias into the reconstructed CMB and/or remove parts of the primary signal. This is particularly true for Galactic foregrounds, whose complex, multi-scale processes are difficult to accurately simulate.

Thus, when ML models have \textit{direct} access to the primary CMB signal during training, the network can learn and inadvertently modify the cosmological signal, or process it along with foregrounds due to imperfect simulations. Within this viewpoint, this bias could propagate through the analysis pipeline and be amplified in downstream inference, leading to systematic errors in cosmological parameter estimation that are difficult to trace or correct. As CMB experiments push towards $\sigma(r)\sim10^{-3}$~\citep{Tristram2022}, ML bias can significantly undermine searches for inflationary $B-$mode polarization. These reasons can make direct ML reconstructions difficult to physically interpret. Therefore, the signal-preserving property is interesting for ML-based CMB component separation. To address this challenge, Ref.~\cite{mccarthy24_ml} presented a novel framework for recovering the CMB with ML that is blind to the cosmological signal of interest. This ensures that the final reconstruction respects the signal-preserving property of ILC. With their ML model,~\cite{mccarthy24_ml} achieved \textit{unbiased} CMB reconstructions with variances that are up to five times lower than that of the ILC solution. While this approach is powerful, it inherits the multi-frequency data requirement of the ILC.

Ref.~\cite{mccarthy24_ml} focused on multifrequency separation. Here, we extend this signal-preserving ML framework for single-frequency reconstruction by leveraging inter-scale correlations within Galactic foregrounds. The key insight is that while \textit{primordial} CMB B-modes should exhibit statistical independence across angular scales (consistent with a {statistically isotropic,} Gaussian random field), Galactic emission processes induce significant correlations between foreground structures at different harmonics \citep{lazarian2000turbulence, cho2002mhd}. This arises from the underlying physics of the interstellar medium and critically distinguishes foregrounds from the primary CMB. As a result, this property enables the use of small-scale information to predict and remove large-scale patterns without multi-frequency data~\citep{Kamionkowski_2014}.

Previous studies have shown that Galactic dust foregrounds exhibit statistical anisotropies that can be uniquely exploited for component separation~\citep{Kamionkowski_2014}. The physical origin of these anisotropies is partly the coherent alignment of dust with the Galactic Magnetic Field (GMF) on few-degree scales, creating preferred polarization directions \citep{SE_2025,Draine_1998}. These large-scale patterns ($\ell<80$) can be estimated and removed using statistical estimators constructed from small-scale ($\ell > 90$) information~\citep{Philcox_2018}. Since this relies purely on geometric information it can be implemented using single frequency data, making it complementary to multi-frequency separation methods. 

In this work, we present a new single-frequency ML approach to extract inter-scale correlations in Galactic dust emission, while adopting the signal-preserving framework introduced in \cite{mccarthy24_ml}. Rather than relying on analytic estimators (as done in~\citep{Philcox_2018}), our method uses convolutional neural networks (CNNs) to learn complex correlations between small- and large-scale features directly from the simulated maps (i.e., at the field-level). Specifically, we investigate how small-scale information ($\ell > 200$) can be used to predict and remove foregrounds at large-scales ($\ell < 200$) where the signal of interest (the primordial B-mode) resides. We emphasize that this is \textit{signal-preserving} by construction since large- and small-scale modes in the primary CMB are statistically independent. 

We demonstrate the effectiveness of this method using the \textsc{DustFilaments} simulations~\citep{Herv_as_Caimapo_2022}, and show that using just small-scale B-modes as input, our reconstructions can achieve high harmonic correlation coefficient averaged over multipoles, $\langle\rho(\ell)\rangle_{\ell}$, relative to the simulated foregrounds at the large scales of interest. {Leveraging the independence between primary $T/E$ and $B$-modes\footnote{Here we again assume $(i)$ Guassian fields and $(ii)$ $C^{\rm{TB}}_{\ell}={C^{\rm{EB}}_{\ell}=0}$ for the primary CMB}, we later augment the CNN with temperature and E-mode polarization}, which further increases $\langle\rho(\ell)\rangle_{\ell}$. Moreover, we explore hybrid approaches that combine multi-frequency cleaning with the inter-scale information. This achieves a reduction in the CMB reconstruction variance by nearly an order of magnitude compared to the standard ILC solution, demonstrating the substantial non-Gaussian and anisotropic information extracted by the CNN. 

This paper is structured as follows. We first present the inter-scale foreground estimation framework in Section~\ref{sec:interscale}. Next, we describe the training configurations {in Section~\ref{sec:architecture_training}  and the simulations used to train the networks in Section~\ref{sec:dustfilaments}}. We then describe the metrics used to evaluate the network performance in Section~\ref{sec:metrics}. In Section~\ref{sec:results}, we present the network predictions and comprehensive evaluations of the resulting foreground and CMB reconstructions. Finally, we conclude in Section~\ref{sec:discussion-conclusion} by discussing the implications, limitations, and future directions of this work.

\section{Signal-Preserving Inter-Scale Foreground Estimation}\label{sec:interscale}

Here, we present the general formalism of signal-free reconstruction with ML, and apply it to the case of inter-scale component separation with the goal of cleaning large-scale B-mode polarization. Inspired by the framework introduced in Ref.~\cite{mccarthy24_ml}, we use the following observation model for multi-frequency CMB B-mode polarization maps:
\begin{equation}
B_i(\hat{\mathbf{n}}) =S(\hat{\mathbf{n}}) + F_i(\hat{\mathbf{n}})\label{eq:general_obs_model}
\end{equation}
where $i$ indicates a frequency channel, $\hat{\mathbf{n}}$ is a unit vector on the observation sphere, $S(\hat{\mathbf{n}})$ is the primordial signal,\footnote{We work in CMB thermodynamic temperature units where $S(\hat{\mathbf{n}})$ has the same amplitude in each frequency channel by definition.} and $F_i(\hat{\mathbf{n}})$ are the foregrounds at frequency $i$, which are uncorrelated with $S(\hat{\mathbf{n}})$. The goal is to design a neural network $\mathcal{N}$ to estimate and remove $F_i(\hat{\mathbf{n}})$ from $B_i(\hat{\mathbf{n}})$. This reconstructs $S(\hat{\mathbf{n}})$ without biasing or removing parts of it, {as the neural network never encounters the cosmological signal}. The key to ensuring that $\mathcal{N}$ obeys this signal-preserving property is two-fold:
\begin{itemize}
    \item \textbf{Signal-free inputs:} First, inputs of $\mathcal{N}$ must be uncorrelated with $S(\hat{\mathbf{n}})$. This ensures the network cannot learn anything about $S(\hat{\mathbf{n}})$ from the inputs and thereby does not have access to information that can potentially bias $S(\hat{\mathbf{n}})$. This includes simulation misspecification and incomplete physical models in the data. However, the inputs should contain information about the components $F_i(\hat{\mathbf{n}})$ that one wishes to remove.
    \item \textbf{Signal-free targets:} Second, the network is constrained to predict targets that are uncorrelated with $S(\hat{\mathbf{n}})$. This enables one to subtract the network prediction from  $B_i(\hat{\mathbf{n}})$ without inadvertently removing parts of  $S(\hat{\mathbf{n}})$. If the network has inherent biases or inaccurate predictions, its predictions would not affect the signal we wish to preserve.
\end{itemize}

This dual constraint ensures that even if the network's learned mapping is imperfect or if simulations contain modeling errors, the primary CMB signal remains unbiased for cosmological inference down the line. A verification of this property is that the cross-power spectrum between the reconstructed signal $\hat{S}$ and the true signal $S$ equals the auto-power spectrum of the true signal~\citep{mccarthy24_ml}:
\begin{equation}
    \langle \hat{S} S \rangle = \langle S S \rangle.
    \label{eq:unbiasedness_condition}
\end{equation}
In this work, we satisfy these criteria by exploiting inter-scale correlations in Galactic dust: while primordial CMB modes at different multipoles are statistically independent, polarized dust couples across scales. We therefore aim to recover the large-scale signal where primordial B-modes are expected, using for foreground cleaning only small-scale information where foregrounds dominate and the primary CMB is negligible even for non-zero $r$. To this end, we {high- and low-pass} filter the maps in harmonic space at $\ell_{\rm{cut}}= 200$ and decompose the observation model as
\begin{align}
B_{L,i}(\hat{\mathbf{n}}) &= S_L(\hat{\mathbf{n}}) + F_{L,i}(\hat{\mathbf{n}}) \label{eq:B_large} \\
B_{S,i}(\hat{\mathbf{n}}) &= S_S(\hat{\mathbf{n}}) + F_{S,i}(\hat{\mathbf{n}}) \label{eq:B_small}
\end{align}
where {subscripts $L/S$ denote maps that retain only large-scale ($\ell < 200$)/small-scale ($\ell > 200$) information.}

\subsection{Inter-Scale Learning: Single-Frequency Inputs}\label{sec:interscale_single_frequency}

We begin with inter-scale reconstruction at a \textit{single frequency}, {which we choose to be at 220 GHz}. In this case, the channel $i$ in the decompositions described in Eqn.~\ref{eq:B_large} and Eqn.~\ref{eq:B_small} can be eliminated.   

\begin{figure}[ht!]
    \centering
    \includegraphics[width=1.0\textwidth]{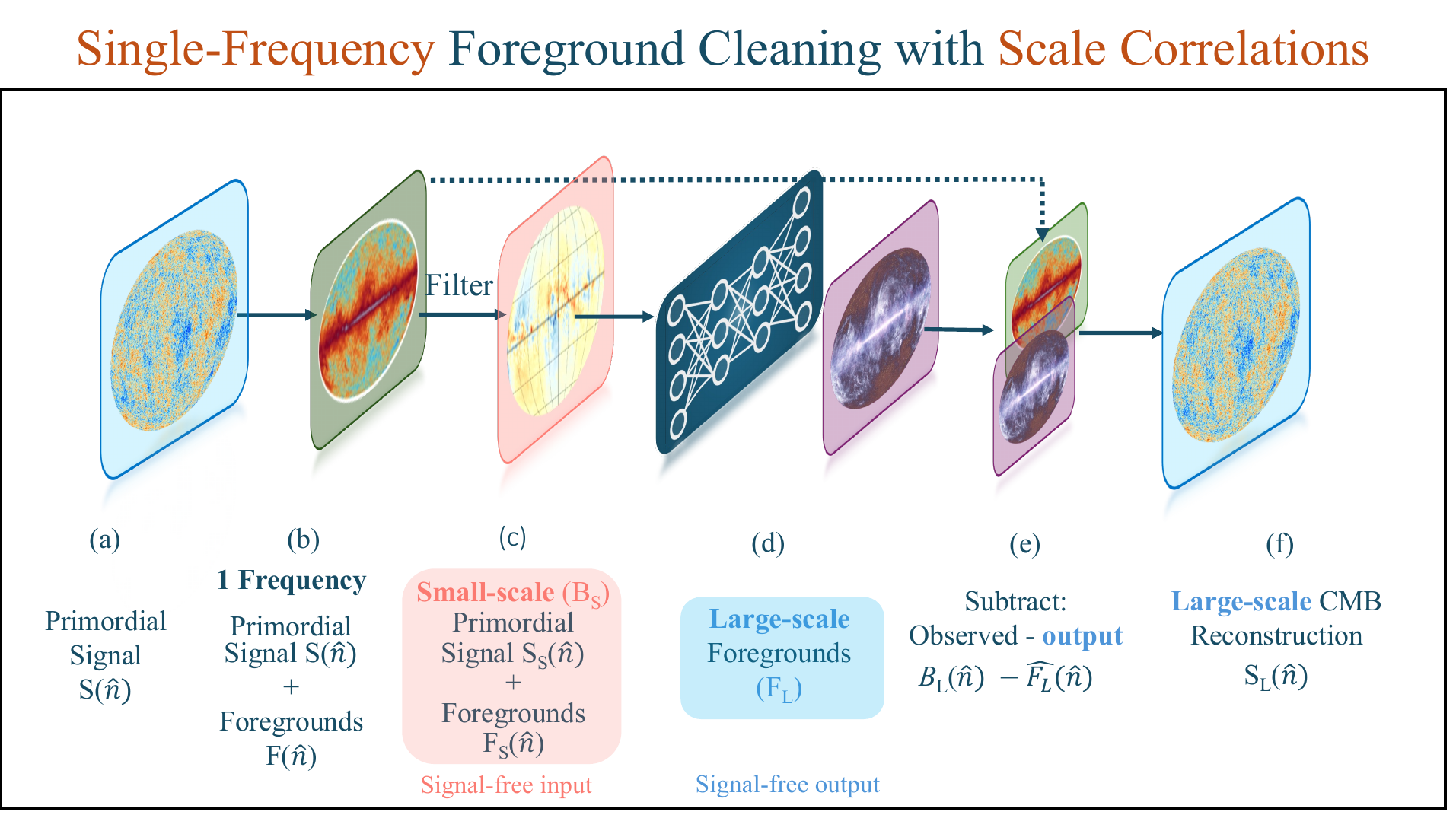}
    \caption{Schematic of our inter-scale machine learning framework for signal-preserving foreground removal. \textbf{(a)} The combination of primary CMB signal and foregrounds at a single frequency gives the observed, contaminated map $B(\hat{\mathbf{n}})$ \textbf{(b)}. Since this map contains the signal of interest, it cannot be directly fed to the network. However, we can decompose it into large-scale ($B_L$, $\ell < 200$) and small-scale ($B_S$, $\ell > 200$) components. \textbf{(c)} The filtered small scale map is then fed to the network as a \textit{signal-free} input. \textbf{(d)} The network is trained to predict the large-scale foreground component $\hat{F}_L$. \textbf{(e)} $\hat{F}_L$ is then subtracted from the observed large-scale map $B_L$ (which contains $S_L +  F_L$) to produce the NN-cleaned map, {$\hat{S}_L(\mathbf{\hat{n}})$}. Since the input $F_S$ is statistically independent of the large-scale CMB signal $S_L$, the prediction $\hat{F}_L$ is also independent of $S_L$, ensuring the cosmological signal is preserved in the final map. }
    \label{fig:interscale-pipeline}
\end{figure}

\subsubsection{Input: Small-scale B-modes}\label{sec:interscale_b}
First, we use $B_{S}(\hat{\mathbf{n}})$ as the signal-free input to reconstruct the large-scale foregrounds, $F_{L}(\hat{\mathbf{n}})$. During training, we remove the small-scale primary CMB component $S_S(\hat{\mathbf{n}})$ from the network input. Given that the CMB is a Gaussian random field, $S_S(\hat{\mathbf{n}})$ behaves as a noise-like contaminant that is statistically independent of both $F_S(\hat{\mathbf{n}})$ and $F_L(\hat{\mathbf{n}})$. Hence, it provides no useful information that could aid the reconstruction, and including it could potentially hinder the network's ability to learn the foreground inter-scale correlations. However, we restore it during model evaluation when we calculate the correlation between the cleaned and {true} CMB. The network prediction, which we denote as $\hat{F}_L(\hat{\mathbf{n}})$, is thus a function of only the small-scale foregrounds:
\begin{equation}
\hat{F}_L(\hat{\mathbf{n}}) = \mathcal{N}(B_S(\hat{\mathbf{n}})) = \mathcal{N}(F_S(\hat{\mathbf{n}})). \label{eq:F_L_prediction}
\end{equation}
Notice that both the input and targets are explicitly uncorrelated with the signal of interest for this reason.

We then later subtract the prediction $\hat{F}_L(\hat{\mathbf{n}})$ from $B_L(\hat{\mathbf{n}})$ to obtain our reconstruction of $S_L(\hat{\mathbf{n}})$:
\begin{equation}
\hat{S}_L(\hat{\mathbf{n}}) = B_L(\hat{\mathbf{n}}) - \hat{F}_L(\hat{\mathbf{n}}) = S_L(\hat{\mathbf{n}}) + F_L(\hat{\mathbf{n}}) - \mathcal{N}(F_S(\hat{\mathbf{n}})) \label{eq:B_L_reconstruction}
\end{equation}
This first network investigates the extent to which correlations exist between $F_S$ and $F_L$, and whether a network can capture these correlations via a non-linear mapping to accurately recover the independent primary CMB signal. Fig.~\ref{fig:interscale-pipeline} is a schematic of this network, illustrating the preservation of large-scale primary CMB B-modes, $S_L(\hat{\mathbf{n}})$, throughout the reconstruction pipeline.

\subsubsection{Inputs: Small-scale B-modes + Temperature + E-Modes}\label{sec:interscale_t_e}
\begin{figure}[ht!]
    \centering
    \includegraphics[width=1.0\textwidth]{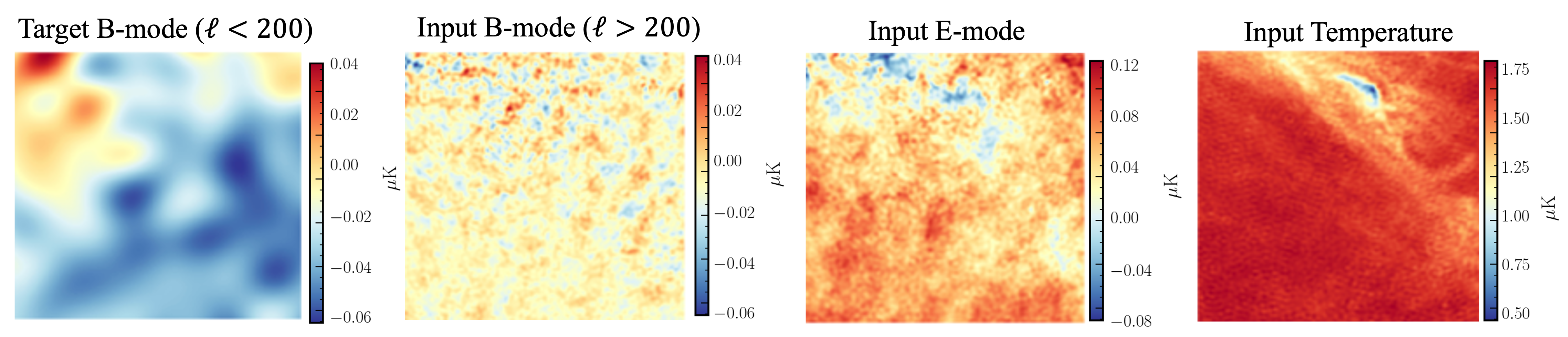}
    \caption{Example training data for the inter-scale foreground prediction objective with T/E mode augmentation. From left to right: (a) {Target patch} containing the large-scale B-mode foreground component ($F_L^B$, $\ell < 200$) (b) Input patch containing small-scale B-mode information ($T_S^B$, $\ell > 200$). (c) Input patch containing the full E-mode polarization map ($E(\hat{\mathbf{n}})$, all $\ell$). (d) Input patch containing the full temperature map ($T(\hat{\mathbf{n}})$, all $\ell$). Panels (b), (c), and (d) are concatenated into a multi-channel input to the network. }
    \label{fig:augmented_training_example}
\end{figure}

We consider a second network in which the inputs are augmented with additional information that is statistically independent of the primary CMB B-modes, but correlated with large-scale Galactic foregrounds, and can therefore improve the reconstruction. Specifically, we include CMB temperature maps $T(\hat{\mathbf{n}})$ and E-mode polarization maps $E(\hat{\mathbf{n}})$ as parallel network input channels. This is primarily motivated by the coupling of E-modes, B-modes, and temperature through the Stokes parameters that quantify the intensity and polarization of the electromagnetic field~\citep{Kosowsky_1996}. However, additional physical processes in the interstellar medium can generate T,E,B correlations. For instance, magnetic misalignment of filament orientation can generate TE and TB correlations~\citep{Planck:2016cww, Vacher_2023}. Other large-scale driving mechanisms in the Galactic dust populations  such as stellar winds and spiral shocks can also induce displacements in the magnetized fluid. Such longitudinal displacements create density, or intensity, structures that correlate with magnetic field structures, thereby coupling E and B modes~\citep{Elmegreen_2004, Kritsuk_2018, 2017ApJ...839...91C, 2023ApJ...946..106C, Clark_2021, Herv_as_Caimapo_2025}. Fig.~\ref{fig:augmented_training_example} shows an example of the three input channels for a training patch for this network, along with the corresponding target. As highlighted by the dashed region, there are visibly correlated structures across the three channels that can be leveraged to inform the network output. The network prediction in this network is:
\begin{equation}
\hat{F}_L^B(\hat{\mathbf{n}}) = \mathcal{N}_{\text{aug}}(B_S(\hat{\mathbf{n}}), T(\hat{\mathbf{n}}), E(\hat{\mathbf{n}})) = \mathcal{N}_{\text{aug}}(F_S^B(\hat{\mathbf{n}}), F^T(\hat{\mathbf{n}}), F^E(\hat{\mathbf{n}})), \label{eq:F_L_aug_prediction}
\end{equation}
{where $F^T, F^E, F^B$ are the temperature, E-mode, and B-mode foregrounds,} and $\mathcal{N}_{\text{aug}}$ is the augmented neural network.

Critically, the augmented inputs are still signal-free since the primary temperature ($S^T$), E-mode ($S^E$), and primordial B-mode ($S^B_{S,\,L}$) components are {independent} at all scales. Hence, we discard the primary CMB components $S^T$ and $S^E$ from the temperature and E-mode inputs during training in Eqn.~\ref{eq:F_L_aug_prediction}, as they act as random noise that is not correlated with the target component nor the component we wish to preserve. Again, we emphasize that we include the primary CMB B-modes when we evaluate the network to compute the correlation between the final CMB reconstruction and the primary CMB.

The final cleaned reconstruction is:
\begin{equation}
\hat{S}_L(\hat{\mathbf{n}}) = B_L(\hat{\mathbf{n}}) - \hat{F}_L^B(\hat{\mathbf{n}}) = S_L(\hat{\mathbf{n}}) + F_L(\hat{\mathbf{n}}) - \mathcal{N}_{\text{aug}}(F_S^B(\hat{\mathbf{n}}), F^T(\hat{\mathbf{n}}), F^E(\hat{\mathbf{n}})) \label{eq:B_L_aug_reconstruction}.
\end{equation}

\subsection{Hybrid: Frequency-difference + Inter-scale Learning}\label{sec:multifreq_interscale}

Next, we explore a hybrid approach that combines inter-scale learning with the multi-frequency framework introduced in~\citep{mccarthy24_ml}. This network leverages the correlation between small-scale and large-scale structures in the {observed} maps, as well as the correlations of the foreground fluctuations between different frequency channels. We use four frequency channels, with the frequencies chosen to match those measured by SO for the purposes of modeling and removing Galactic emission~\citep{SimonsObservatory2019}: 95, 145, 225 and 280 GHz. Note that since the \textsc{DustFilaments} code does not provide simulations at 225 and 280 GHz, we use 220 and 270 GHz as proxies. 

To incorporate multi-frequency channels, we first compute the spatial ILC reconstruction of the $B$-mode map, which is the linear combination of the observed maps, $B_i(\mathbf{\hat{n}})$, using the ILC weights $w_j$ optimized to achieve minimal variance of the reconstruction, $\hat{B}^{\rm{ILC}}$:
\begin{align}
    \hat{B}^{\rm{ILC}} =  \sum_j w_j B_j(\hat{\mathbf{n}}) = S(\mathbf{\hat{n}}) + \sum_j w_j F_j(\hat{\mathbf{n}}) , \quad \sum_i a_i w_i = 1, \label{eq:ilc_weights}
\end{align}
where $w_i$ are the spatial ILC's optimized weights and $a_i$ are the frequency-dependent mixing coefficients that scale the components' SEDs with respect to the CMB thermodynamic temperature (for the primary CMB, $a_i = 1$ for all frequencies)~\citep{eriksen2004ilc}. Note that the sum constraint on $w_j$ is what ensures that $S(\mathbf{\hat{n}})$ is fully preserved and can thus be pulled out of the sum.

Next, we adopt the choice of inputs from~\citep{mccarthy24_ml}; explicitly, we use the difference between the observed map at each frequency and the ILC CMB reconstruction:
\begin{align}
    \hat{F}^{\rm{ILC}}_i = B_i(\hat{\mathbf{n}}) - \hat{B}^{\mathrm{ILC}}(\hat{\mathbf{n}}) &= F_i(\hat{\mathbf{n}}) - \sum_j w_j F_j(\hat{\mathbf{n}})
\end{align}
{Here, $\hat{F}^{\rm{ILC}}_i$ can be seen as {imperfect foreground estimates} derived using only linear correlations in $F_i$ across the frequencies.} Notice that $\hat{F}^{\rm{ILC}}_i$ is uncorrelated with the signal of interest, which is fully contained in both $B_i$ and $\hat{B}^{\mathrm{ILC}}(\hat{\mathbf{n}})$ and thus canceled out upon subtraction.

To build signal-free outputs, we again follow~\citep{mccarthy24_ml} and train the network to predict signal-free outputs. We choose these to be the ILC residuals, which are defined as:
\begin{equation}
    \hat{{\Delta F}^{\mathrm{ILC}}}(\hat{\mathbf{n}}) = \hat{B}^{\mathrm{ILC}}(\hat{\mathbf{n}}) - S(\hat{\mathbf{n}}) =  \sum_j w_j F_j(\hat{\mathbf{n}})
    \label{eq:ilc_residuals},
\end{equation}
again using the fact that $\hat{B}^{\mathrm{ILC}}$ fully contains the primary CMB signal. Consequently, the signal of interest is removed upon subtraction from the observed map, $B_i$, resulting in a signal-free target. Using this as the network target enables prediction of anisotropic foregrounds \textit{missed} by the ILC, which can then be subtracted from the ILC solution to achieve an ultimate NN-improved reconstruction: 
\begin{equation}
\begin{aligned}
 B^{\mathrm{pred}}_L(\hat{\mathbf{n}}) &=  B^{\mathrm{ILC}}_L(\hat{\mathbf{n}}) - \hat{\Delta F^{\mathrm {ILC}}}_L(\hat{\mathbf{n}}) = \sum _i w_i B_i(\hat{\mathbf{n}}) - \mathcal{N}_{\text{aug}}(\hat{F}^{\mathrm{ILC}}_i(\hat{\mathbf{n}}), B_S(\hat{\mathbf{n}}))\\
&= S_L(\hat{\mathbf{n}}) +  \sum_i w_i F_i(\hat{\mathbf{n}}) - \mathcal{N}\big(\hat{F}^{\mathrm{ILC}}_i(\hat{\mathbf{n}}), B_S(\hat{\mathbf{n}})\big).
\end{aligned}
\label{eq:ultimate_reconstruction}
\end{equation}

We first investigate the combination of frequency-difference maps with small-scale B-modes as inputs to train a network to predict the ILC residuals. This is analogous to the inter-scale learning framework described in Sec~\ref{sec:interscale_b}. Hence, the network prediction can be written as:

\begin{equation}
    \hat{\Delta F^{\mathrm {ILC}}}_L = \mathcal{N}(B_S, \hat{F}^{\rm{ILC}}_i).
    \label{eq:multifreq_interscale_pred}
\end{equation}
Later, we also explore the combination of frequency-difference maps with small-scale B-modes, temperature, and E-modes (at multiple frequencies) as inputs to train a network ($\mathcal{N}$) to predict the ILC residuals. In this case, the network prediction can be written as:
\begin{equation}
    \hat{\Delta F^{\mathrm {ILC}}}_L = \mathcal{N}(B_{S,i}, T_i, E_i, \hat{F}^{\rm{ILC}}_i).
    \label{eq:multifreq_interscale_pred_teb}
\end{equation}
In the subsequent section, we detail the network architectures used to implement both the single-frequency inter-scale network and the hybrid multi-frequency and inter-scale network. 

\section{Network Architecture and Training Procedure}\label{sec:architecture_training}
We employ a UNet architecture \citep{ronneberger2015unet}, which is well-suited for image-to-image prediction tasks that require multi-scale feature learning. The encoder-decoder structure with skip connections also enables the network to capture correlations between small-scale input features (e.g., $\ell > 200$ foreground structures) and large-scale target features (e.g., $\ell < 200$ foreground contamination), while preserving fine-grained spatial details throughout the network depth.

We implement multiple networks to explore different strategies for foreground removal. The input channels for each are specified below:
\begin{itemize}
    \item \textbf{Single-frequency networks}: 1 channel ($B_S$-only) or 3 channels ($T+E+B_S$) at a single frequency
    \item \textbf{Multi-frequency networks}: 8 channels (4-frequency $\hat{F}^{\text{ILC}}_i$ + 4-frequency $B_{S, \, i}$) or 16 channels (previous 8 channels + 4-frequency $E_i$ + 4-frequency $T_i$)
\end{itemize}
The first two networks (1 and 3 input channels, respectively) test inter-scale learning using single-frequency maps decomposed by angular scale. The latter two networks (8 and 16 input channels, respectively) combine frequency-difference maps (from ILC foreground predictions) with inter-scale information. This hybrid approach leverages both frequency-dependent foreground properties and scale-dependent correlations, potentially providing more {information for} foreground removal.

All networks share the same core architecture described in Appendix~\ref{app:nn_archi} with a single-channel output that predicts either large-scale foregrounds (for single-frequency networks) or ILC residuals (for multi-frequency networks).

Moreover, all networks are trained to minimize the mean squared error (MSE) between predicted and true targets:
\be
L_{\text{MSE}} = \mathbb{E} \left\| \hat{Y} - Y \right\|^2,
\ee
where $\mathbb{E}$ is the average over training patches, $\hat{Y}$ is the network prediction and $Y$ is the target (large-scale foregrounds for inter-scale networks, ILC residuals for frequency-difference networks). We use the \textsc{Adam} optimizer~\citep{kingma2017} during training, with learning rate and weight decay tuned via the Bayesian hyperparameter selection wrapper \textsc{Optuna}~\citep{optuna_2019}, and early stopping based on validation loss. Complete architecture specifications, including layer initializations, hyperparameters, and training configurations, are provided in Appendix~\ref{app:training_config}.

\section{Galactic Dust Foreground Simulations with DustFilaments}\label{sec:dustfilaments}

We train and validate our models using simulated full-sky maps from the \textsc{DustFilaments} model \citep{Herv_as_Caimapo_2022}, which simulates Galactic thermal dust emission by populating the Galaxy with millions of individual filaments aligned with the GMF. The \textsc{DustFilaments} code\footnote{Available at \url{https://github.com/chervias/DustFilaments}} generates thermal dust maps by integrating a population of 3D filaments painted onto the celestial sphere, reproducing the statistical properties of \Planck 353 GHz dust polarization maps, including angular power spectra and non-Gaussian features in the Minkowski functionals, while providing independent realizations of the dust sky.

We generated 150 random full-sky \textsc{DustFilaments} realizations with HEALPix~\citep{gorski1999healpixprimer, Gorski_2005} resolution $N_{\text{side}} = 1024$ (pixel resolution $\sim 3.4$ arcmin), at frequencies 95, 145, 220, and 270 GHz (matching the SO frequency bands)~\citep{SimonsObservatory2019}. We later refer to these maps as [$B_{\text{95}}, B_{\text{145}}, B_{\text{220}}, B_{\text{270}}$]. A complete description of all simulation parameters used in this work, including filament properties, magnetic field configuration, and spectral energy distribution parameters, can be found in Table 1 of \citep{Herv_as_Caimapo_2022}. Additional details on the simulation normalization procedure and multi-frequency map generation are provided in Appendix~\ref{app:simulation_normalization}. 

Note that the \textsc{DustFilaments} code provides two options for simulating the Galactic plane. The first uses \textit{Planck}’s generalized needlet ILC GNILC dust template constructed from the low-frequency instrument (LFI) and high-frequency instrument (HFI) \textit{Planck} DR2 maps~\citep{planck_2016} instead of populating filaments in the plane with their model. The second option is to isotropically sample filament centers in this region so that sightlines through the Galactic plane are populated in the same manner as all other directions. Effectively, the simulation does not impose an explicit Galactic-plane geometry, but the filaments in this region are still evolved with the magnetic-field model. Since the first option does not simulate small-scale emission in the Galactic plane~\citep{Herv_as_Caimapo_2022}, and since this region is never realistically used or sufficiently cleaned in CMB polarization experiments due to overwhelming foreground contamination, we use the second option for our simulations.

We also generate realizations of primordial CMB B-modes (without lensing) using \textsc{CAMB} \citep{2011ascl.soft02026L} with best-fit \Planck 2018 cosmological parameters~\citep{planck2018params} (see column 2 of Table 1 in ~\citep{planck2018params}) and a tensor-to-scalar ratio of $r=0.008$, yielding $S^B(\hat{\mathbf{n}})$ maps that are statistically independent from the foreground simulations. Observed maps are then constructed as $B(\hat{\mathbf{n}}) = S^B(\hat{\mathbf{n}}) + F^B(\hat{\mathbf{n}})$ at each frequency, combining the CMB signal with the dust foreground contamination. We do this also for $T(\hat{\mathbf{n}})$ and $E(\hat{\mathbf{n}})$.

We extract $128 \times 128$ pixel patches (corresponding to $\sim 7.3^\circ \times 7.3^\circ$ at $N_{\text{side}} = 1024$) from full-sky maps using a Cylindrical Equal-Area (CAR) projection.\footnote{We thank Will Coulton for sharing code to perform this operation.} {We use a cosine window function to apodize the patches with a taper width of 5 pixels on each side.} Approximately $\sim1,280$ patches are extracted per full-sky simulation, yielding a total dataset of $\sim 192,000$ patches. These patches are divided into training (80\%), validation (10\%), and test (10\%) sets, with stratification by simulation index to ensure that patches from the same simulation do not appear in multiple splits to maintain statistical independence between the different datasets.

Finally, we apply a filter in spherical harmonic space to decompose foreground maps into large-scale ($\ell < 200$) and small-scale ($\ell > 200$) components:
\begin{itemize}
    \item \textbf{Large-scale foregrounds} ($F_L$): Target for single-frequency inter-scale networks, representing contamination at the angular scales where primordial B-modes are expected ($\ell \lesssim 200$)
    \item \textbf{Small-scale foregrounds} ($F_S$): Input for inter-scale networks, where foregrounds dominate the signal and the primordial CMB B-mode power is negligible ($\ell > 200$)
\end{itemize}

This scale separation exploits the statistical independence of CMB modes across different multipoles while leveraging the coherent structure of Galactic dust foregrounds across scales. We have tested the robustness of this scale separation by varying the multipole threshold and find that our results are not significantly affected by changes in the scale cut. Nevertheless, this choice has not been optimized to maximize the information available to the neural network, and a systematic exploration of optimal angular-scale selection for B-mode information extraction is left for future work.

\section{Validation Metrics}\label{sec:metrics}

We evaluate model performance using the following validation metrics, with the harmonic correlation and residual foreground power as our primary figures of merit for reconstruction quality and cleaning efficacy.

\begin{enumerate}
    \item We assess correlations in harmonic space using a pseudo-spectrum estimator computed on each CMB patch. We work in the flat-sky limit and compute binned 2D Fourier cross-spectra $C_{\ell}^{F_L\times \hat{F_L}}$ for the apodized patches, $F_L \, \mathrm{and} \, \hat{F_L}$ at 220 GHz. We use bin widths $\Delta \ell = 50$. 

    We divide the cross-power spectrum by the geometric mean of the auto-power spectra to yield the normalized harmonic correlation coefficient $\rho(\ell)$,
        \begin{equation}
        \rho(\ell) = \frac{C_\ell^{\hat{F}_L \times F_L}}{\sqrt{C_\ell^{\hat{F}_L \times \hat{F}_L} C_\ell^{F_L \times F_L}}}.
        \end{equation}
    We report the average correlation coefficient over large-scale multipoles ($\ell < 200$): $\langle\rho(\ell)\rangle_{\ell}$. We also report the mean harmonic correlation across all test patches, along with the 16th and 84th percentiles (two-tailed equivalent to $\pm 1\sigma$ for a Gaussian distribution) of the distribution across patches at each multipole bin, providing uncertainty bands that reflect the patch-to-patch variation in reconstruction quality.

    \item To quantify how well we remove foregrounds, we report the residual foreground power remaining after subtracting the predicted foregrounds from the observed patch,
    \begin{equation}
        f_{\rm{resid}} = \big\langle \bigl(1-[\rho(\ell)]^2\bigr)\big\rangle_{\ell<200},
    \end{equation}
    which gives the fraction of large-scale foreground power remaining after removal.

    \item We also compute the mean squared error between the network prediction $\hat{Y}$ and the corresponding target $Y$,
    \begin{equation}
    \text{MSE}_{\text{pred}} = \frac{1}{N_p} \sum_{p=1}^{N_p} \left(\hat{Y}(\hat{\mathbf{n}}_p) - Y(\hat{\mathbf{n}}_p)\right)^2,
    \label{eq:pred_mse_metric}
    \end{equation}
    where $Y$ denotes either the large-scale foreground map ($F_L$) for single-frequency networks, or the ILC residual ($\Delta F^{\mathrm{ILC}}$) for multi-frequency and hybrid networks, and $\hat{Y}$ is the corresponding network prediction. For an unbiased estimator the MSE measures the variance of the reconstructed CMB signal, which is the quantity minimized by ILC-based methods. However, for non-Gaussian fields, variance minimization does not guarantee minimal reconstruction error, since dust polarization anisotropies depend on higher-order moments not captured by the ILC. We therefore compute the MSE on a patch-by-patch basis for the various networks discussed in Section~\ref{sec:interscale} as a complementary metric.

    \item Finally, we evaluate spatial alignment using the Pearson correlation coefficient,
        \begin{equation}
        \rho_{\text{spatial}} = \frac{\mathrm{Cov}(\hat{X}, X)}{\sigma_{\hat{X}} \, \sigma_X},
        \end{equation}
    where $X$ denotes either the foreground map ($F_L$) or the primary CMB signal ($S$), and $\hat{X}$ its corresponding prediction. The covariance is normalized by the standard deviations of the predicted and true fields, $\sigma_{\hat{X}} \, \sigma_X$. 
    To assess the statistical significance of these correlations, we compute the distribution of \textit{null} correlations between the prediction and all other, non-corresponding patches in the test dataset. The prediction's actual correlation is then compared against this null distribution. Correlations significantly exceeding the null distribution (e.g., $> 2\sigma$) confirm that the network has learned predictive spatial patterns rather than correlating by chance.
\end{enumerate}

\section{Results}\label{sec:results}
\subsection{Single-Frequency Inputs}\label{sec:results_interscale}
\subsubsection{Small-Scale B-modes only}
\begin{figure}[ht!]
    \centering
    \includegraphics[width=1\linewidth]{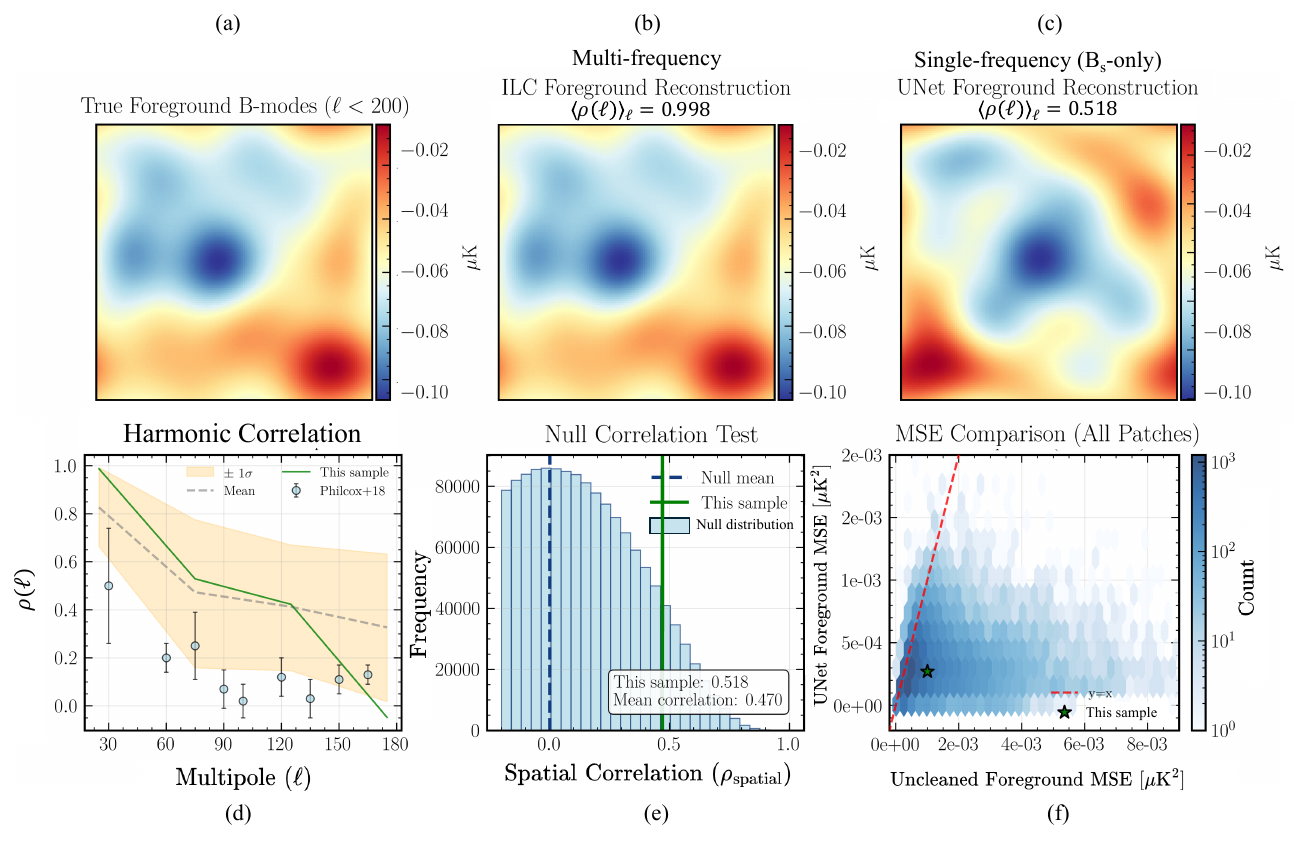}
    \caption{Foreground prediction quality for a representative test patch. Top row: (a) Large-scale foreground B-modes ($\ell < 200$, target), (b) Large-scale ILC foreground estimate {($\hat{F}^{\rm{ILC}}_{\rm{220}}$)} computed using all \textit{four frequency} channels and shown for comparison, (c) UNet foreground prediction. The spatial correlations achieved by both methods relative to the true foregrounds are indicated above the respective panel. Bottom row: (d) Harmonic correlation coefficient comparing UNet and observed CMB reconstructions to pure CMB with mean across the test set denoted by the dashed line and the  shaded bands denoting $\pm 1\sigma$ errors. The mean harmonic correlation across all test patches and multipole bins is $\langle\rho(\ell)\rangle_{\ell} = 0.51 \pm 0.31$. However, the correlation is larger at low-$\ell$, with $\rho(\ell<80) \simeq 0.5$--$0.8$. (e) Null correlation test histogram showing the spatial correlation for this patch ($\rho_{\rm{spatial}}=0.52$, green dashed line) compared to the distribution of correlations obtained when pairing the prediction with all other target patches from different realizations (histogram). The actual correlation exceeds the null distribution mean (approximately zero, dashed vertical line) by more than $2\sigma$ for all test patches, indicating that the reconstructions are statistically significant rather than chance correlations (see Fig.~\ref{fig:correlation_comparison} for the distribution of matched prediction-target correlations across the test set). (f) MSE comparison across all test patches with this sample's metric indicated. The mean MSE for UNet foreground predictions is $3.44 \times 10^{-4} \mu \mathrm{K}^2$, while the single-frequency ILC (no-cleaning) MSE is $1.50 \times 10^{-3} \mu \mathrm{K}^2$. 
    \label{fig:b_only_fg_recon}}
\end{figure}
We first investigate how well the network can predict large-scale B-mode {polarization foregrounds} using only small-scale information at a single frequency (220 GHz). This minimal network establishes a baseline performance and tests whether inter-scale correlations exist in B-mode foregrounds alone. Fig.~\ref{fig:b_only_fg_recon} shows an example of the UNet foreground prediction for a test patch. The top row features: (a) the large-scale foreground B-modes ($\ell < 200$, target); (b) the large-scale ILC foreground prediction ($\hat{F}^{\rm{ILC}}$ at 220 GHz){, which is computed via a multi-frequency ILC with all four frequency channels}; and (c) the UNet prediction. As it can be seen, the ILC reconstruction is much more accurate than the UNet prediction, with the former exhibiting a spatial correlation of approximately unity while the latter only achieves $\rho_{\rm{spatial}} \sim 50\%$. However, note that the ILC has the additional advantage of using multi-frequency information, while the UNet is restricted to a single frequency. 

A fairer comparison is obtained by evaluating the ILC in the single-frequency limit, where it has access to the same information as the network and its reconstruction is equivalent to the uncleaned (observed) map, $B_{\rm 220}$. We define this as the baseline MSE. In this limit, the UNet is expected to outperform the ILC, which is indeed seen in the scatterplot (f). As can be seen in the figure, the average MSE between the UNet predicted and true large-scale foregrounds is $3.44 \times 10^{-4}\mu \mathrm{K}^2$ on the test set, compared to the baseline MSE, which is $\sim$4 times larger at $1.50 \times 10^{-3} \mu \mathrm{K}^2$ Moreover, over $99\%$ of the test patches exhibit a reconstruction MSE lower than the baseline MSE, aligning with the expectation that the UNet-cleaned map should not have larger error than the uncleaned foregrounds.

In the bottom row of {Fig.}~\ref{fig:b_only_fg_recon}, we show the validation metrics for this reconstruction sample and how it compares with the performance on all test patches. First, panel (d) shows the harmonic correlation between the UNet reconstruction and the truth (green solid line), plotted on top of the mean $\pm 1\sigma$ error band across all patches (grey dashed line with shaded region). The mean harmonic correlation across all test patches and multipole bins is $\langle\rho(\ell)\rangle_{\ell} = 0.51 \pm 0.31$. Furthermore, the harmonic correlation shows scale-dependence, decreasing from 0.83 $\pm$ 0.29 at $\ell\sim 25$ to $0.33 \pm 0.3$ at $\ell\sim175$.

\begin{table}[ht!]
\centering
\caption{Performance metrics for single-frequency (220 GHz) ILC versus UNets trained to predict large-scale ($\ell < 200$) foregrounds. Note that in this limit, the ILC solution is simply the uncleaned (observed) map, containing the primary CMB and foregrounds. The MSE with respect to the primary CMB is hence just the mean square values of the pixels of the foregrounds (first row). The $B_S$-only network is trained using only small-scale ($\ell > 200$) B-modes. For the ILC variants (first two rows), the spatial and harmonic correlation coefficients are computed between the ILC CMB reconstruction and true primary B-modes, while for the UNets (rows 3-7) they are between the predicted and true foregrounds. It can be seen that all UNets perform better than the baseline (MSE ~4–7$\times$ lower), as expected. E-modes are more predictive than B-modes (0.68 vs 0.46 correlation). As expected, using additional components (T,E,B) improves performance -- the maximal network, trained on $T+E+B_S$, achieves $\langle\rho(\ell)\rangle_{\ell}=0.79$ correlation.}

\label{tab:single_freq_performance}
\small
\begin{tabular}{ccccc}
\textbf{Single-Frequency} & \textbf{Harmonic Corr.} & \textbf{Residual Power} & \textbf{MSE} & \textbf{Spatial Corr.} \\
\textbf{Cleaning Method} & $\langle\rho(\ell)\rangle_{\ell}$ & $f_{\rm{resid}}$ & [$\mu \mathrm{K}^2$] & $\rho_{\text{spatial}}$ \\
\midrule
No Foreground Cleaning & 0.02 & 0.999 & $1.50 \times 10^{-3}$ & 0.08 \\
Spatial ILC (3 channels: $B, T, E$) & 0.03 & 0.999 & $5.22 \times 10^{-3}$ & 0.11 \\
$B_S$-only & $\mathbf{0.51}$ & $\mathbf{0.704}$ & $\mathbf{3.44 \times 10^{-4}}$ & $0.46$ \\
$T$-only & $0.46$ & $0.734$ & $3.47 \times 10^{-4}$ & $0.44$ \\
$E$-only & $0.71$ & $0.481$ & $2.53 \times 10^{-4}$ & $0.68$ \\
$T+E$ & $0.74$ & $0.449$ & $2.35 \times 10^{-4}$ & $0.71$ \\
$T+E+B_S$ & $\mathbf{0.79}$ & $\mathbf{0.376}$ & $\mathbf{2.09 \times 10^{-4}}$ & $0.76$ \\
\bottomrule
\end{tabular}
\end{table}

It is also interesting to compare our UNet reconstructions with the results of \citep{Philcox_2018}, {where the authors} developed a complementary single-frequency approach using an estimator of the hexadecapole anisotropy in the B-modes of Galactic dust emission. While our methods differ---{in~\citep{Philcox_2018} statistical patterns  in the position-dependent 2D power spectrum were exploited}, whereas our approach learns {more general} inter-scale correlations through ML---we both tackle single-frequency foreground prediction using {non-Gaussian} information. Moreover, in \citep{Philcox_2018} {the estimator was applied} to a $1\%$ sky region (see Fig. 8 of \citep{Philcox_2018}) compared to our training patches which cover $\sim0.13\%$ of the sky. They also use the \citet{Vansyngel_2018} simulations which implement different physical models of polarized Galactic foregrounds than \textsc{DustFilaments}. Despite these differences, we include their results in panel (d) as a valuable benchmark comparison since both approaches utilize single-frequency information for estimating large-scale foregrounds. Visually inspecting Fig.~9 of~\citep{Philcox_2018}, the hexadecapole estimator attains an average correlation coefficient of $\sim 15\%$ across $\ell<300$, which is substantially lower than our UNet's $\langle\rho(\ell)\rangle_{\ell}\sim 0.51$. This lower correlation is expected, since their method extracts only fourth-order angular patterns (the $\cos{4\phi}$ and $\sin{4\phi}$ modulation of the 2D power spectrum), whereas our UNet utilizes field-level information from all small-scale modes at $\ell > 200$. This comparison highlights the ability of neural networks to learn and capture inter-scale correlations beyond those accessible to anisotropy templates based on fixed angular decompositions.

Next, we examine the correlation in real space between the reconstruction and the truth for the same test patch. This is seen in panel (e) of Fig.~\ref{fig:b_only_fg_recon}. The vertical green line at $\rho_{\rm{spatial}}=0.52$ denotes this patch's correlation. Across the test set, we achieve a mean spatial correlation of $\rho_{\text{spatial}} = 0.46 \pm 0.25$ (mean $\pm$ standard deviation) between predicted and true large-scale foregrounds, which is consistent with the harmonic correlation results. Moreover, we demonstrate the null correlation test by comparing the spatial correlation of this patch against the distribution of correlations between the prediction and every other target patch in the test set, which are displayed in the histogram. We find that the reconstruction correlations lie $>2\sigma$ above the mean of the null distribution (which is approximately zero) for all test patches, confirming statistical significance as opposed to random correlation. 

Overall, inter-scale correlations exist in B-mode foregrounds and can be exploited for large-scale foreground prediction, as demonstrated by $\langle\rho(\ell)\rangle_{\ell}\simeq 0.51$ and a $\sim4.36\times$ MSE improvement over baseline. However, performance is highly variable across patches, with some regions showing strong correlations ($> 0.9$) while others fail ($< 0.2$).\footnote{Since the simulations are statistically homogeneous across the Galactic plane, this variability reflects random fluctuations in the foreground realizations rather than systematic differences between different regions of the sky.} This variability, combined with the moderate mean correlation, indicates that single-frequency B-mode information alone provides insufficient information for robust reconstruction across all sky regions.

These limitations motivate the investigation of the next section, where we explore whether augmenting the input with complementary information from temperature and E-mode polarization can improve reconstruction performance and reduce variability across patches.

\subsubsection{Temperature and E-Modes}
\begin{figure}
    \centering
    \includegraphics[width=1\linewidth]{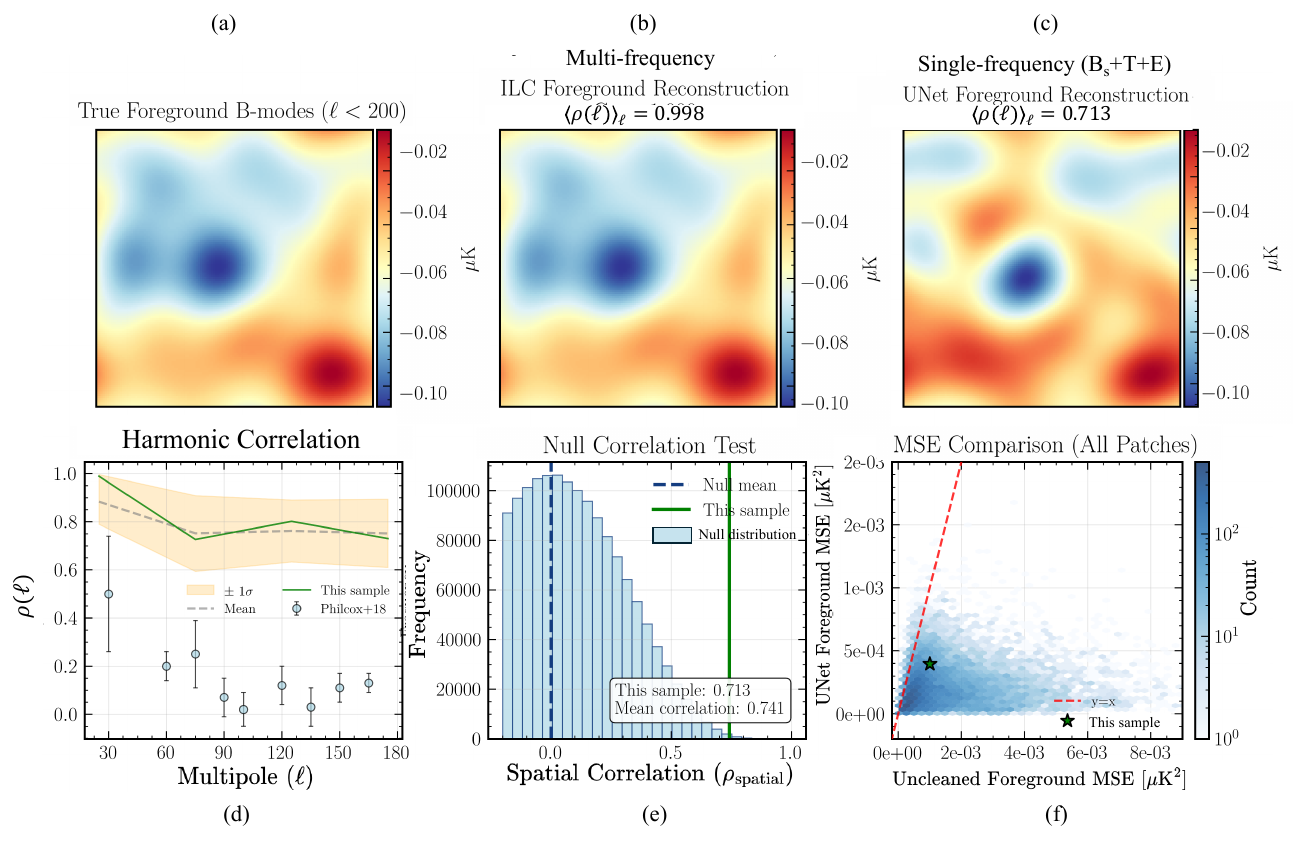}
    \caption{Foreground prediction from the UNet trained on $T+E+B_S$ for the same patch displayed in Fig.~\ref{fig:b_only_fg_recon}, with the same layout. Compared to the $B_S$-only network, this network more accurately {recovers the polarized foreground morphology, with a MSE that is $\sim1.64\times$ smaller}. The inclusion of additional inputs also leads to substantially higher spatial and harmonic correlation coefficients when averaged over all test patches (see Table~\ref{tab:single_freq_performance}). Notably, panel (d) illustrates that the harmonic correlation coefficient remains high across all large-scale multipoles, with $\langle\rho(\ell)\rangle_{\ell}\simeq 0.79$ and lower uncertainty (as indicated by the yellow shaded region).}
    \label{fig:teb_fg_recon}
\end{figure}

\begin{figure}[ht!]
    \centering
    \begin{minipage}[b]{0.49\textwidth}
        \centering
        \includegraphics[width=\linewidth]{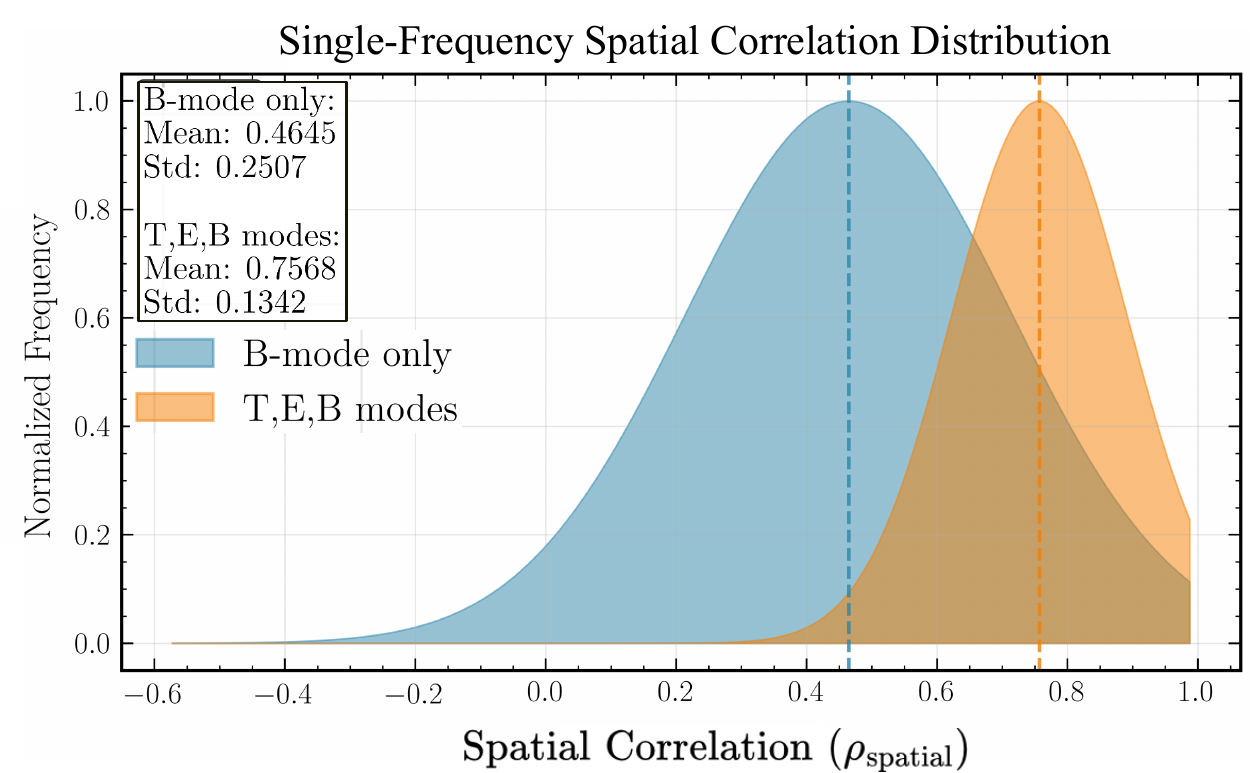}
    \end{minipage}
    \hfill
    \begin{minipage}[b]{0.49\textwidth}
        \centering
        \includegraphics[width=\linewidth]{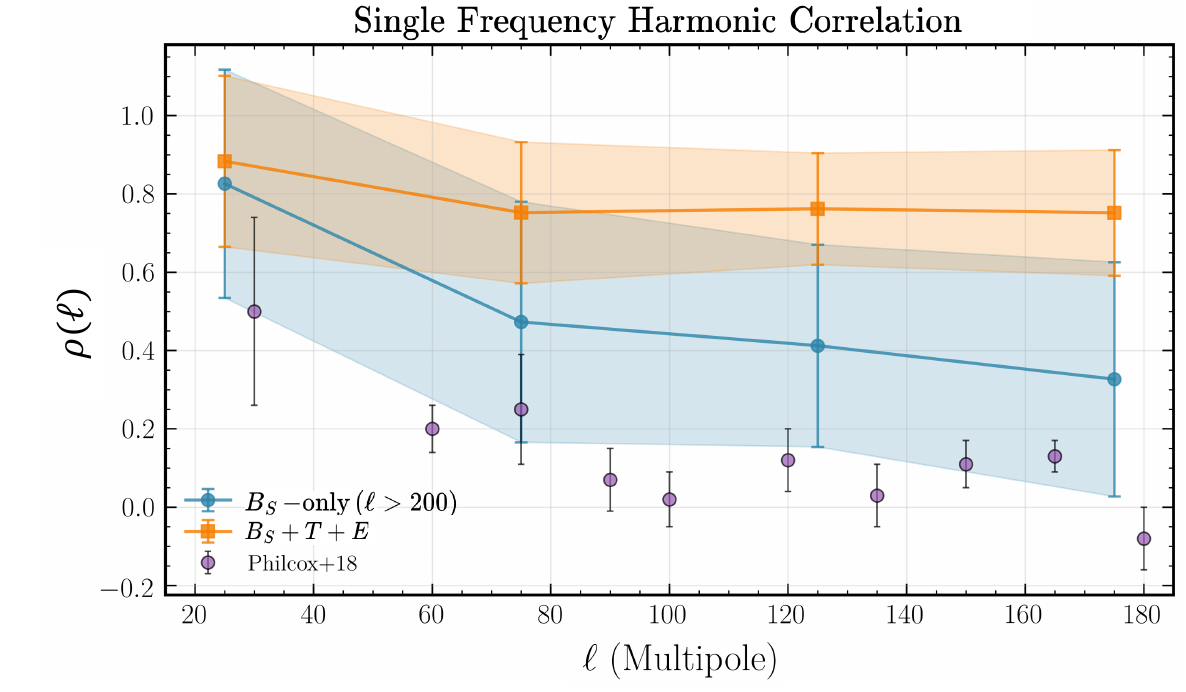}
    \end{minipage}
    \caption{Comparison of spatial and harmonic correlations for reconstructions from $B_S$-only and $T+E+B_S$ networks. \textbf{Left:} Distribution of spatial correlations between predicted and true foregrounds for $B_S$-only (blue) and $T+E+B_S$ (orange) networks. Distributions are approximated from mean and standard deviation statistics using Gaussian distributions. Vertical dashed lines mark the mean correlation for each network. \textbf{Right:} Harmonic correlation coefficient. Both metrics demonstrate that the $T+E+B_S$ network substantially outperforms $B_S$-only, with higher correlations and lower spread.}
\label{fig:correlation_comparison}
\end{figure}
The limited performance of the UNet trained with $B_S$-only input motivates exploring whether additional CMB components can provide complementary information to improve foreground prediction. Physically, we expect temperature and E-mode polarization to contain relevant information about $B$-mode foreground structure, as discussed in Section~\ref{sec:interscale}. While cross-correlations between $B$ and either $T$ or $E$ vanish by parity invariance for the primary CMB~\citep[e.g.,][]{Kamionkowski_1997}, this parity invariance need not hold for foreground emission, such as Galactic dust, where physical processes such as misalignment can couple the modes \citep{2017ApJ...839...91C, Kandel_2017, Elmegreen_2004, Tegmark_2000, 2023ApJ...946..106C}. Thus, E-mode measurements can supply additional information about B-mode foreground morphology, potentially encoding features of the polarization field that are not captured by small-scale B modes alone. Fig.~\ref{fig:teb_fg_recon} shows the improved foreground estimation for the same test patch when the UNet is trained on $T+E+B_S$ inputs rather than $B_S$ alone.

Fig.~\ref{fig:correlation_comparison} compares test-set averaged spatial and harmonic correlations between UNet predictions and the corresponding truth for the $B_S$-only and $T+E+B_S$ networks. The left panel shows the distribution of spatial correlations, averaging at $\rho_{\text{spatial}} = 0.76 \pm 0.13$ across the test set. This is a $\sim 63\%$ improvement over the $B_S$-only network ($0.46 \pm 0.25$). The significant reduction in standard deviation from $0.25$ to $0.13$ also indicates more consistent performance across patches. Similarly, the right panel compares the harmonic correlations, where the $T+E+B_S$ UNet achieves an average of $\langle\rho(\ell)\rangle_{\ell} = 0.79$, compared to $\langle\rho(\ell)\rangle_{\ell} = 0.51$ for $B_S$-only. The harmonic correlation shows consistent improvement across all multipole bins, and with reduced uncertainty compared to $B_S$-only at all scales. The MSE progression from baseline to multi-component networks, as listed in Table~\ref{tab:single_freq_performance}, highlights the benefit of incorporating multiple components. Overall, all three metrics demonstrate that the additional information from $T+E+B_S$ substantially improves foreground predictions. 

\begin{figure}[ht]
    \centering
    \includegraphics[width=1\linewidth]{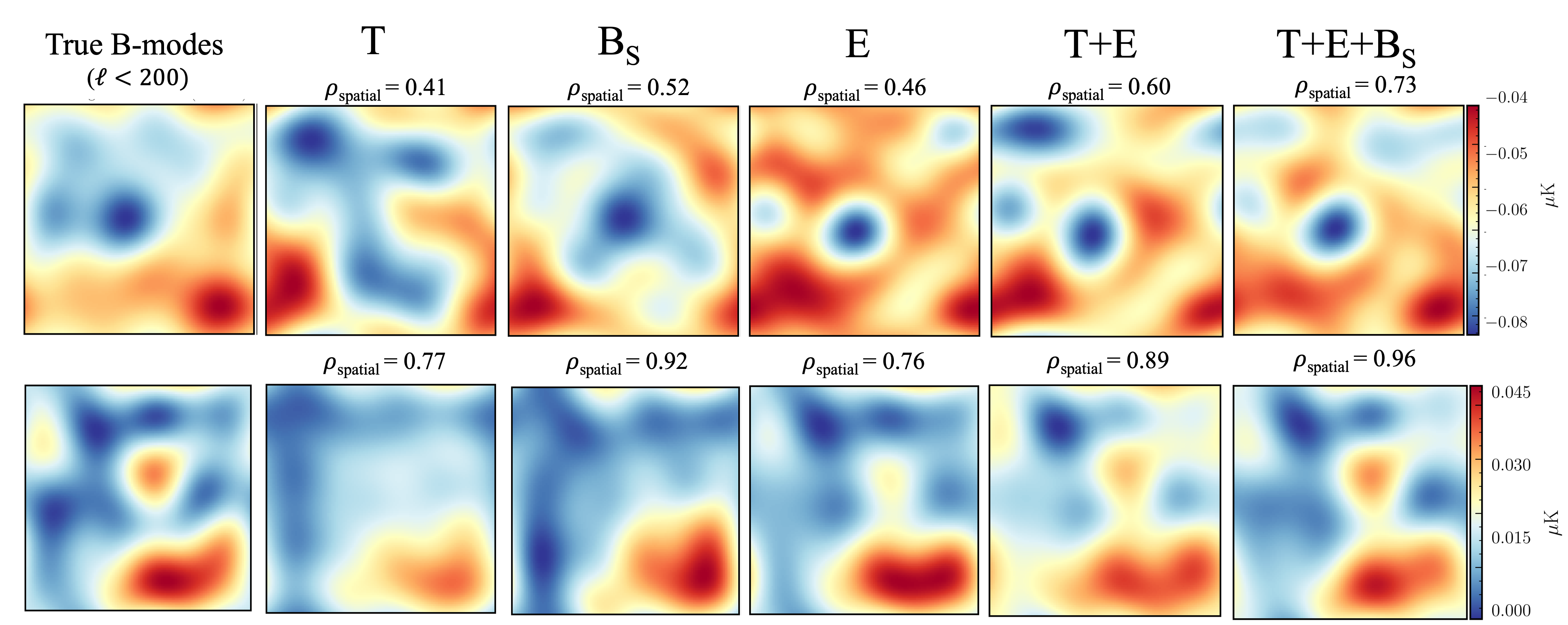}
    \caption{Each row corresponds to a different test-set target patch (large-scale B-modes), followed by foreground predictions from different networks. From left to right, the inputs are temperature ($T$), small-scale B-modes only ($B_S$), E-modes ($E$), $T+E$, and $T+E+B_S$. The spatial correlation coefficient with the true large-scale B-mode foreground is shown above each reconstructed patch. For some regions (e.g., the bottom row), E-mode information provides the dominant improvement in reconstruction accuracy. In other cases, E modes alone are insufficient and yield accurate reconstructions only when combined with additional inputs. Moreover, T-only networks perform poorly in recovering B-mode foreground structure.}
    \label{fig:rows_single_freq}
\end{figure}

\begin{figure}
    \centering
    \includegraphics[width=1.0\linewidth]{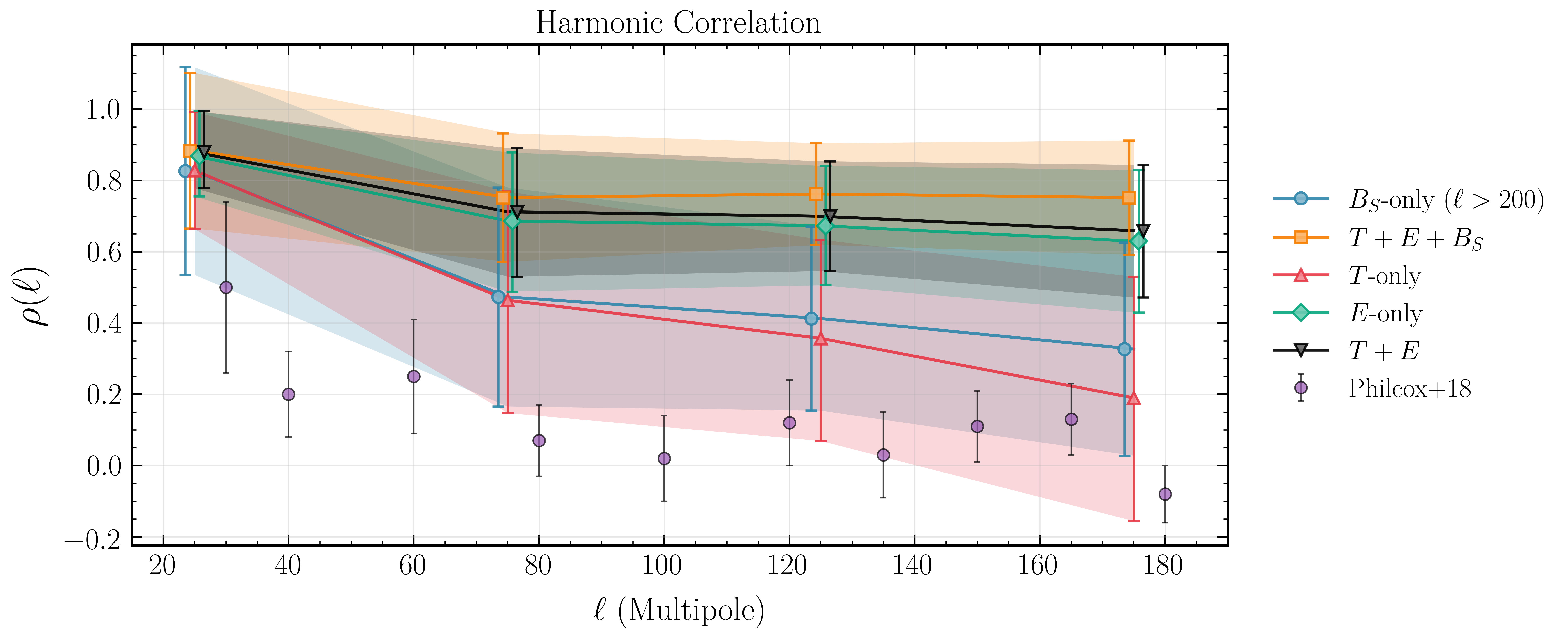}
    \caption{The harmonic correlation coefficient as a function of multipole for the large-scale ($\ell<200$) foreground predictions achieved by the UNets using various \textit{single-frequency} networks. The solid lines denote the mean results for the test set patches while the shaded regions indicate $\pm 1\sigma$ uncertainty. As can be seen, the network trained on $T+E+B_S$ input attains the highest correlations across all scales, with significantly reduced uncertainties. Moreover, the E-modes (green) provide substantially more information than the small-scale B-modes (blue) for large-scale B-mode reconstruction. For comparison, we also include the results obtained by the B-mode hexadecapolar estimator of \citep{Philcox_2018} (see text for discussion of methodological and dataset differences).\footnote{$\rho(\ell)$ and errorbars estimated from Fig.9 of~\citep{Philcox_2018}} We have shifted the UNet points in the x-axis for visual clarity. 
    }
    \label{fig:single_freq_compare_corr}
\end{figure}

To understand the relative contributions of each foreground component, we compare reconstructions for different networks. Table~\ref{tab:single_freq_performance} summarizes the performance metrics for all single-frequency networks. Fig.~\ref{fig:rows_single_freq} visualizes this comparison for foreground estimations achieved on two test patches (one per row), with each row showing the target large-scale foregrounds (first column) and the predictions obtained from networks trained on $T$, $B_S$, $E$, $T+E$, and $T+E+B_S$ inputs. Fig.~\ref{fig:single_freq_compare_corr} summarizes the corresponding harmonic correlations as a function of multipole for all networks. The mean spatial correlations of the reconstructions across the various networks are: $T+E+B_S$ ($\rho_{\text{spatial}} = 0.76 \pm 0.13$) $>$ $T+E$-only ($\rho_{\text{spatial}} = 0.71 \pm 0.15$) $>$ $E$-only ($\rho_{\text{spatial}} = 0.68 \pm 0.17$) $>$ $B_S$-only ($\rho_{\text{spatial}} = 0.46 \pm 0.25$) $>$ $T$-only ($\rho_{\text{spatial}} = 0.44 \pm 0.26$). This is also supported by their harmonic correlations, as shown in the right panel of Fig.~\ref{fig:correlation_comparison}. These results demonstrate several key insights: (1) E-modes are most informative of B-mode foregrounds ($\langle\rho(\ell)\rangle_{\ell} = 0.71$), substantially outperforming both small-scale $B_S$-only ($\langle\rho(\ell)\rangle_{\ell} = 0.51$) and T-only ($\langle\rho(\ell)\rangle_{\ell} = 0.46$). (2) Combining T and E-modes ($\langle\rho(\ell)\rangle_{\ell} = 0.74$) provides additional benefit beyond E-only. (3) Adding small-scale B-modes to T+E ($\langle\rho(\ell)\rangle_{\ell} = 0.79$) yields the best reconstructions. We speculate that E-only performs better than $B_S$-only because the E-modes provides access to information on all angular scales (not just small scales), which is more helpful for large-scale foreground prediction. The stronger performance of E-only compared to T-only is also expected since polarization is more informative for B-mode foregrounds than intensity. Physically, this also reflects how E-modes respond to density fluctuations correlated with the Galactic magnetic field~\citep{2023ApJ...946..106C}. The additional improvement from including small-scale B-modes in the maximal $T+E+B_S$ network indicates that inter-scale correlations within B-modes themselves also contribute beyond what can be inferred from cross-component information.

In summary, augmenting the $B_S$-only training with temperature and E-modes provides substantial complementary information for large-scale foreground prediction. The $T+E+B_S$ network achieves a $63\%$ improvement in mean spatial correlation and a $55\%$ improvement in $\langle\rho(\ell)\rangle_{\ell}$ compared to $B_S$-only, while simultaneously reducing the correlation uncertainties by $\sim47\%$. The MSE is also reduced by a factor of $\sim7.18$ compared to the ILC baseline (vs a factor of $\sim.36$ for $B_S$-only). Although these results are preliminary, they suggest that single-frequency observations with multi-component $T+E+B_S$ inputs can substantially improve large-scale foreground prediction in the idealized case of noiseless CMB measurements. However, future tests are needed to assess whether this can improve CMB reconstructions under more realistic settings and achieve tighter cosmological constraints. Moreover, as we know from the strong performance of ILC methods, multi-frequency correlations provide substantial information for foreground removal. We therefore explore a hybrid approach that combines multi-frequency and inter-scale correlations in the next section.

\begin{table}
\centering
\caption{Performance metrics for multi-frequency inter-scale hybrid UNets. All networks are trained to predict ILC residuals $\Delta F^{\rm{ILC}}$ using frequency-difference maps and inter-scale information. Using $\hat{F}^{\text{ILC}}_i$-only inputs achieves a $\sim$2$\times$ reduction in CMB reconstruction MSE compared to the 4-channel spatial ILC, and reduces the residual foreground power $f_{\rm{resid}}$ by a similar factor. This demonstrates that the UNet has learned non-Gaussian and anisotropic foreground information that the ILC was unable to mitigate. Adding small-scale B-modes ($\hat{F}_i + B_S$) reduces the MSE by $\sim$2.7$\times$ over the 4-channel UNet, demonstrating the advantage of inter-scale correlations (see Fig.~\ref{fig:multi-freq-NN}). Furthermore, augmenting with temperature and E-modes ($\hat{F}_i + T + E + B_S$ , 16-channel network) results in an additional $\sim$1.5$\times$ reduction -- decreasing the MSE by nearly an order of magnitude compared to the 4-channel spatial ILC. The spatial and harmonic correlation coefficients for all UNets indicate that multi-frequency information is already able to achieve high-fidelity reconstructions. Yet, the progressive improvement in the correlations with additional inputs of T, E, and small-scale B-modes demonstrate that they are indeed useful for foreground prediction rather than redundant. Interestingly, adding $T, E$ to the spatial ILC worsens the MSE and correlation metrics due to higher foreground variance.\footnote{We have checked that the variance of the ILC solution has not increased with more input channels, as expected for the ILC's variance minimization algorithm}}
\label{tab:multifreq_performance}
\small
\begin{tabular}{ccccc}
\toprule
\textbf{Multi-Frequency} & \textbf{Harmonic Corr.} & \textbf{Residual Power} & \textbf{MSE} & \textbf{Spatial Corr.} \\
\textbf{Method} & $\langle\rho(\ell)\rangle_{\ell}$ & $f_{\rm{resid}}$ & $\text{MSE}_{\text{CMB}}$ & $\rho_{\text{spatial}}$ \\
\midrule
Spatial ILC (4-channel: $B$) & $0.99864$ & $2.69\times10^{-3}$ & $3.42 \times 10^{-6}$ & $0.9983$ \\
Spatial ILC (8-channel: $B+T$) & $0.99360$ & $1.22\times10^{-2}$ & $1.37 \times 10^{-5} $ & $0.9962$ \\
Spatial ILC (8-channel: $B+E$) & $0.99594$ & $7.69\times10^{-3}$ & $7.47 \times 10^{-6}$ & $0.9964$ \\
Spatial ILC (12-channel: $B+T+E$) & $0.99101$ & $1.68\times10^{-2}$ & $1.86 \times 10^{-5}$ & $0.9944$ \\
$\hat{F}^{\text{ILC}}_i$-only (4-channel) & $0.99950$ & $\mathbf{1.00\times10^{-3}}$ & $1.51 \times 10^{-6}$ & $0.9985$ \\
$\hat{F}^{\text{ILC}}_i + B_S$ (8-channel) & $0.99976$ & $\mathbf{4.71\times10^{-4}}$ & $5.61 \times 10^{-7}$ & $0.9993$ \\
$\hat{F}^{\text{ILC}}_i + B_S + T + E$ (16-channel) & $0.99982$ & $\mathbf{3.62\times10^{-4}}$ & $3.66 \times 10^{-7}$ & $0.9996$ \\
\bottomrule
\end{tabular}
\end{table}
\subsection{Frequency-Difference + Inter-Scale UNet}\label{sec:results_combined}

\subsubsection{Frequency Difference + Small-Scale B}\label{sec:multifreq_interscale_b}
\begin{figure}[ht!]
    \centering
    \includegraphics[width=1\linewidth]{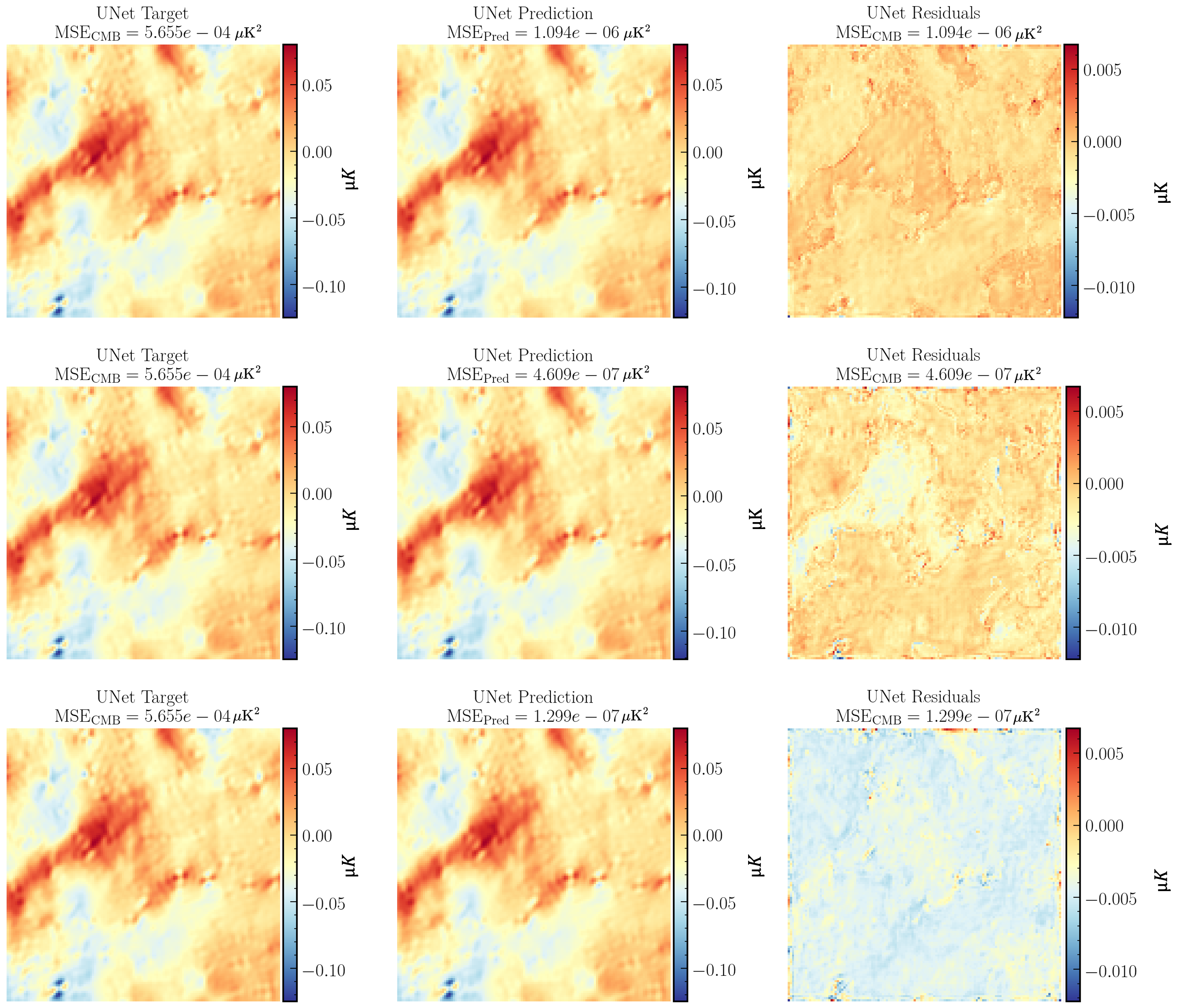}
    \caption{Similar to Fig. 12 in Ref.~\cite{mccarthy24_ml}, showing the target ILC residuals (first column), the UNet prediction (second column), and the final residual of the UNet-corrected reconstruction, which is obtained after removing the UNet prediction from the ILC solution (third column). As it can be seen in the first columns, the ILC residuals are anisotropic. These remnant structures are what ILC cannot remove and reflect the need for non-linear reconstruction. \textbf{Top row:} UNet trained on just $\hat{F_i}$. While producing the highest residuals of the three networks, it successfully mitigates anisotropic ILC structures. \textbf{Middle row:} UNet trained on $\hat{F_i} + B_S$, achieving a substantially lower MSE compared to that of the ILC residual. \textbf{Bottom row:} UNet trained on $\hat{F_i} + B_S + T + E$. As it can be seen, this network is able to remove almost all of the foreground variance that the ILC solution still contained, reducing the residual MSE by over $\mathcal{O}(10^3)\times$ (from $\mathrm{MSE} = 5.65 \times 10^{-4}\mu \mathrm{K}^2$ to $\mathrm{MSE}_{\text{CMB}} = 1.29 \times 10^{-7}\mu \mathrm{K}^2$). Overall, increasing network inputs progressively eliminates residual foreground structures in the ILC solution, consistently suggesting that the additional information with each network is useful for predicting anisotropic foregrounds.
}
    \label{fig:compare-all}
\end{figure}

While the previous inter-scale results demonstrate that foreground prediction is feasible using only a \textit{single frequency}, they do not leverage the rich information available in the SEDs of foreground components across multiple frequencies. Notably, Ref.~\cite{mccarthy24_ml} used that ILC foreground estimates, which are constructed by subtracting the ILC estimate from each frequency channel, provide signal-free inputs that exploit the frequency-dependent correlations of foregrounds while preserving the CMB signal. Specifically, the ILC foreground estimates $\hat{F}^{\rm{ILC}}_i = B_i - \hat{S}^{\mathrm{ILC}}$ capture foreground information through their SED variations, which are distinct from the scale-dependent correlations exploited by inter-scale methods. Hence, they should provide complementary constraints on foreground morphology. We therefore investigate a hybrid approach that combines frequency-difference maps with small-scale B-modes, and then later temperature and E-modes as well. We discuss these results below.

First, we train a UNet on only ILC foreground estimates $\hat{F}^{\rm{ILC}}_i$ at each frequency for comparison with the later augmented networks. We find that this UNet achieves a MSE of $1.51 \times 10^{-6}\mu \mathrm{K}^2$ which is $\sim$2$\times$ lower than the baseline ILC solution ($3.42 \times 10^{-6}\mu \mathrm{K}^2$). The correlations in harmonic ($\langle\rho(\ell)\rangle_{\ell}$) and pixel space ($\rho_{\rm{spatial}}$) are also improved despite the already high correlation achieved by standard ILC. These results agree with those of \citep{mccarthy24_ml} (despite being trained on and applied to very different simulations than those of~\citep{mccarthy24_ml}, where the PySM~\cite{pysm3} simulations were used) and demonstrate that the UNet is able to effectively identify and remove the anisotropies that remain in foreground residuals (Fig.~\ref{fig:compare-all}). 

While this multi-frequency approach is already performing very well, we are motivated to ask how much improvement can be attained with additional information. Hence, we augment the inputs with the small-scale B-mode maps, $B_S$, and later temperature and E-modes. By simultaneously exploiting both spectral correlations across frequencies and structural correlations across angular scales, we hypothesize that these networks can potentially achieve higher fidelity reconstructions compared to either approach alone.
\begin{figure}[ht!]
    \centering
    \includegraphics[width=0.8\linewidth]{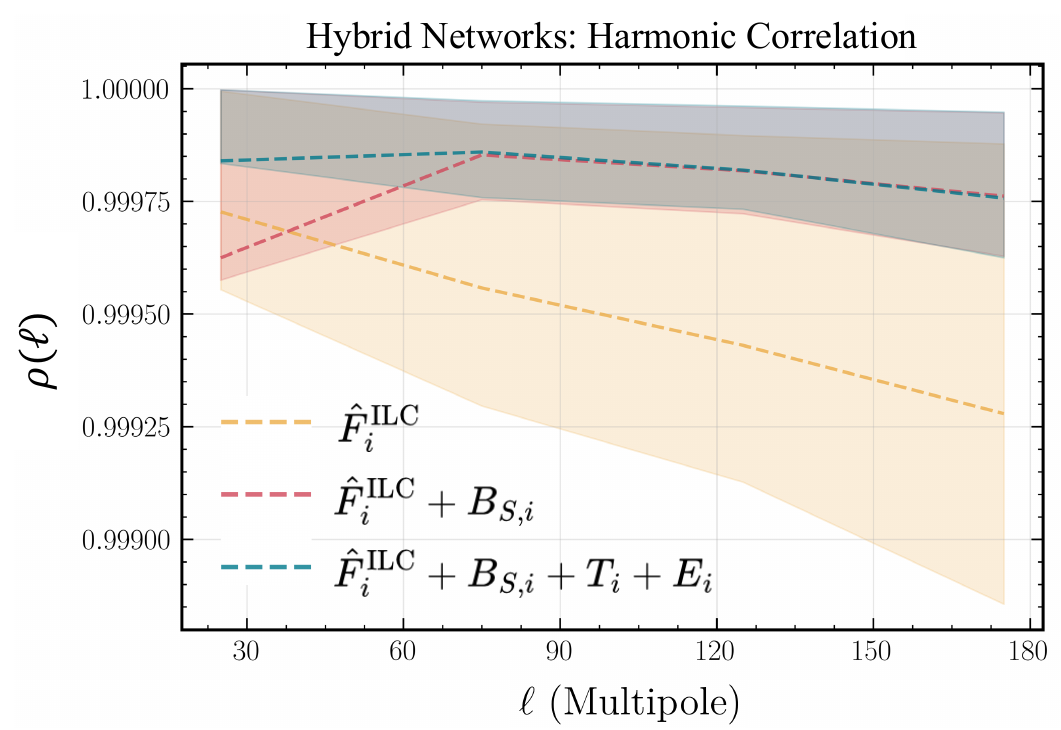}
    \caption{Harmonic correlation achieved by the hybrid multi-frequency and inter-scale networks. The dashed lines illustrate the mean correlation coefficients across the test set and the shaded regions indicate the $\pm 1\sigma$ spread. When using a combination of multi-frequency and $B_S$ inputs (red), the UNet achieves $\langle\rho(\ell)\rangle_{\ell} = 0.99976$ ($f_{\rm{resid}} = 4.71 \times 10^{-4}$), compared to $\langle\rho(\ell)\rangle_{\ell} = 0.99950$ ($f_{\rm{resid}} = 1.00 \times 10^{-3}$) for the $\hat{F}^{\rm{ILC}}_i$-only UNet (yellow). This demonstrates the impact of inter-scale information for foreground estimation. Augmenting with $T+E$ inputs (blue) further tightens the correlation uncertainty at the largest scales, and increases the mean harmonic correlation to $\langle\rho(\ell)\rangle_{\ell} = 0.99982$ ($f_{\rm{resid}} = 3.62 \times 10^{-4}$).}
    \label{fig:multi-freq-NN}
\end{figure}

We first train a UNet on a combination of the frequency-difference maps and small-scale B-modes, as described in Section~\ref{sec:multifreq_interscale}. This UNet receives as input the ILC foreground estimates $\hat{F}^{\rm{ILC}}_i$ at each of the four frequency channels (93, 145, 220, and 280 GHz) along with the corresponding small-scale B-mode maps $B_S$ at each frequency, resulting in 8 parallel inputs (Eq.~\ref{eq:multifreq_interscale_pred}). Similar to the $\hat{F}^{\rm{ILC}}_i$-only UNet, this network is trained to predict the ILC residual $\Delta F^{\rm{ILC}}$, which is then subtracted from the ILC solution to produce an improved CMB reconstruction (Eqn.~\ref{eq:ultimate_reconstruction}).

The middle row of Fig.~\ref{fig:compare-all} illustrates this UNet's reconstruction on a test patch, which can be compared to that of the $\hat{F}^{\rm{ILC}}_i$-only UNet in the first row. Each row shows the ILC residual (target $\Delta F^{\rm{ILC}}$), the UNet prediction ($\hat{\Delta F^{\rm{ILC}}}$), and the final residual after correction ($\Delta F^{\rm{ILC}} - \hat{\Delta F}^{\rm{ILC}}$). As can be seen in the first column, the ILC residuals are strongly anisotropic, making them difficult for the ILC to remove. While both UNets successfully learn these anisotropies, the network trained with $B_S$ removes them more effectively, especially targeting the anisotropic structure near the center of the patch {for this sample}. This comparison emphasizes the advantage of jointly using inter-scale correlations with multi-frequency data. The MSE values displayed above the first and third columns are computed with respect to the CMB and quantify the variance of the reconstruction obtained by the ILC and UNet, respectively. With $B_S$ inputs, the UNet achieves a residual variance that is substantially lower compared to the ILC solution for this patch. Averaged across all test patches, this UNet achieves $\mathrm{MSE}_{\text{CMB}} = 5.61 \times 10^{-7}\mu \mathrm{K}^2$, which is $\sim 2.7\times$ smaller than that of the $\hat{F}^{\rm{ILC}}_i$-only UNet and $\sim 6\times$ smaller than the 4-channel spatial ILC ($3.42 \times 10^{-6} \mu \mathrm{K}^2$).

This demonstrates that adding small-scale B-mode information provides substantial complementary information beyond what can be extracted from frequency correlations alone. Evidently, inter-scale correlations are able to capture structures in the foreground maps that are not fully encoded in the frequency-dependent SED variations, allowing the network to better predict the spatial distribution of residual foregrounds. Similarly, using only inter-scale B-mode information is insufficient, as the single-frequency $B_S$-only network discussed in Section~\ref{sec:results_interscale} produced reconstructions that are three orders of magnitude larger than the hybrid UNet (see Table~\ref{tab:single_freq_performance}). This demonstrates that frequency-difference and inter-scale information are not redundant as both are necessary towards optimal foreground removal. 

Motivated by the substantial improvement that temperature and E-mode information provided in the single-frequency inter-scale networks (Section~\ref{sec:results_interscale}), we next investigate whether incorporating T and E modes across all frequency channels can similarly further enhance the multi-frequency hybrid approach.

\subsubsection{Frequency Difference + Small-Scale B + T, E}\label{sec:multifreq_interscale_teb}
\begin{figure}
    \centering
    \includegraphics[width=1\linewidth]{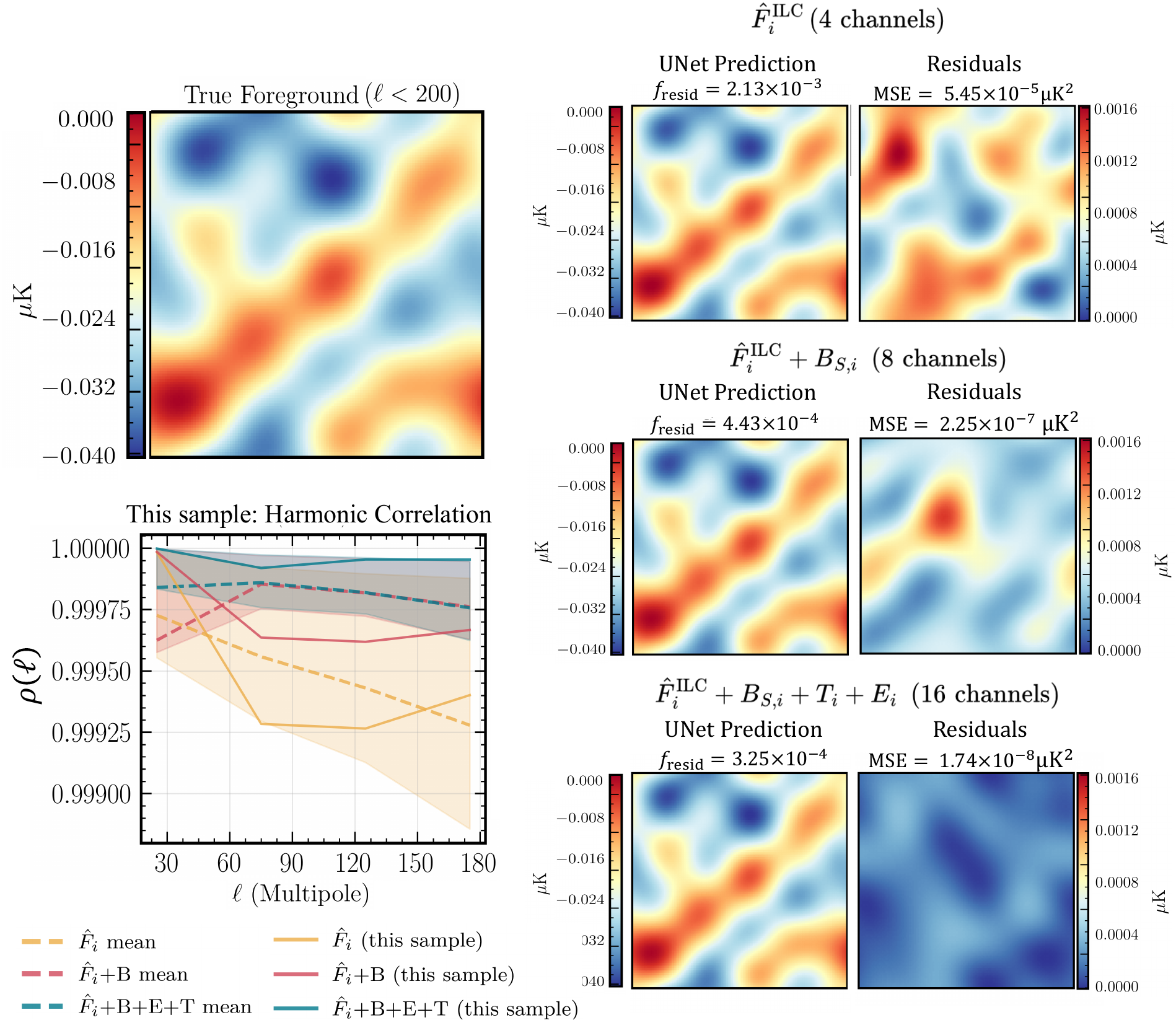}
    \caption{Test patch showing foreground prediction results for the multi-frequency hybrid networks and the corresponding harmonic correlations. Note that the UNet foreground predictions are computed by subtracting the UNet residuals ($\Delta_{\rm{ILC}} $) from the true foregrounds and filtering to retain only the large-scales ($\ell<200$). In the two columns on the right, each row shows the foreground prediction achieved by the respective UNet, accompanied by the residuals with respect to the true foregrounds: $|\hat{F}_{\rm{220}} - F_{\rm{220}}|$. Notably, the residuals decrease as additional input information is incorporated. Although already small for the $\hat{F}_i$-only UNet, their amplitudes are further suppressed with additional inputs, indicating that both inter-scale B-mode correlations and foreground temperature and E-mode information contribute to improved estimation of large-scale B-modes.}
    \label{fig:sample_16ch}
\end{figure}

We now extend the 8-channel network to include temperature and E-mode polarization information across all four frequency channels, resulting in 16-channel inputs. This UNet combines multi-frequency data (frequency-difference maps $\hat{F}^{\rm{ILC}}_i$), inter-scale correlations (small-scale B-modes $B_S$), and cross-component correlations (E-modes $E$ and temperature maps $T$) at each of the four frequencies (Eqn.~\ref{eq:multifreq_interscale_pred_teb}). The bottom row of Fig.~\ref{fig:compare-all} shows the performance of this UNet on the same test patch, demonstrating further reduction in ILC residuals compared to the above two UNets. This 16-channel network achieves a CMB reconstruction MSE of $3.66 \times 10^{-7}\mu \mathrm{K}^2$, representing a $1.53\times$ improvement over the 8-channel network. While this improvement is more modest than the gain from adding inter-scale B-mode information to frequency-difference maps (going from first to second row), it still demonstrates that temperature and E-modes can provide useful complementary constraints on B-mode morphology even when frequency and scale correlations are already used. Moreover, when compared to the UNet trained only on $\hat{F}^{\text{ILC}}_i$, the improvement is substantial, achieving a $4.12\times$ reduction in MSE. In terms of residual foreground power, this network achieves $f_{\rm{resid}} = 3.62 \times 10^{-4}$ across all test patches, which is $2.76\times$ lower than that of the $\hat{F}^{\text{ILC}}_i$ network and $\sim 7\times$ lower than the 4-channel spatial ILC ($f_{\rm{resid}} = 2.69\times10^{-3}$), demonstrating that this maximal network approaches strong CMB recovery with minimal residual foreground contamination.

Fig.~\ref{fig:multi-freq-NN} compares $\rho(\ell \gtrsim 30)$ across all three UNets. As it can be seen, adding $B_S$ to the UNet trained on $\hat{F}_i$-only significantly increases $\rho(\ell \gtrsim 30)$, with improvements being more substantial for smaller scales at $\ell \gtrsim 75$. Moreover, there is a significant reduction in the uncertainty of $\rho(\ell)$. When we further add the temperature and E-modes, it can be seen that there is not much gain on the small scales where the correlation saturates from the $\hat{F}_i+B_S$ inputs.

Fig.~\ref{fig:sample_16ch} shows another example of the estimated large-scale foregrounds for one test patch achieved by these hybrid networks. The two columns on the right compare the UNet foreground estimates and their absolute residuals:~$|\hat{F}_{\rm 220}-F_{\rm 220}|$ for the $\hat{F}_i$-only, $\hat{F}_i+B_S$, and $\hat{F}_i+B_S+T+E$ networks from top to bottom. In each case, the displayed foreground estimate is obtained from the corresponding UNet prediction of the ILC residual and filtered to retain only large scales ($\ell<200$). The corresponding $\rho(\ell)$ for this particular test patch is plotted on the left. As additional inputs are included, the foreground residuals significantly shrink and the harmonic correlations approach unity, consistent with the progressive improvements summarized in Table~\ref{tab:multifreq_performance}. 

\section{Discussion and Conclusions}\label{sec:discussion-conclusion}

In this work, we explored single-frequency CMB reconstruction of Galactic dust foregrounds by extending upon the methods developed in \citep{Kamionkowski_2014, Philcox_2018} and \citep{mccarthy24_ml}. Unlike conventional ILC or foreground template methods, single-frequency CMB component estimation does not rely on spectral correlations or models of the foreground SEDs. This is especially useful for sources with spatially varying SEDs, non-Gaussian structures, and anisotropc statistics -- all of which are prevalent in thermal Galactic dust emission but are not perfectly modeled in simulations nor effectively removed by standard ILC techniques. 

However, whereas \citep{ Philcox_2018} constructed statistical estimators using dust polarization hexadecapole anisotropy pattern, we train CNNs to learn information from all small-scale modes ($\ell > 200$) at the field level to predict large-scale foregrounds. ML is more capable of capturing complex inter-scale correlations and learning the non-trivial Galactic physics that couple different scales and components of dust polarization. 

Importantly, cautioned against the uninterpretability of neural networks and their risk of biasing the cosmological signal, we followed the framework introduced in \citep{mccarthy24_ml} to build a signal-preserving neural network that is blind to the cosmological signal of interest. With the goal of improving constraints on inflation parameters, we identify this signal as the large-scale primary B-mode polarization at $\ell<200$. As a result, any data that has been filtered to contain only $\ell>200$ is considered as signal-free and can be safely fed to the neural network. This ensures that the CMB reconstructions are not biased by the training or reconstruction pipeline, either due to model misspecfication, simulation imperfections, or uninterpretable black-box operations. To quantify residual contamination after cleaning, we use $f_{\rm{resid}} = \langle 1-[\rho(\ell)]^2\rangle_{\ell<200}$, which measures the fraction of large-scale foreground power remaining after subtracting the predicted foregrounds

Using this framework, we first investigated single-frequency reconstruction with various UNets:

\begin{enumerate}
    \item We trained a UNet on small-scale B modes ($\ell>200$) to estimate large-scale B modes ($\ell<200$). This network is able to achieve statistically significant correlations between the reconstruction and the truth, in both harmonic and pixel space (see Table.~\ref{tab:single_freq_performance}). However, it still leaves a residual foreground power of $f_{\rm{resid}}\simeq 0.704$, which is much higher compared to standard spatial ILC that uses multi-frequency information.
    \item Next, we augmented the inputs with additional (signal-free) temperature and E-mode ($T+E$) components at all scales. These allow the UNet to leverage physical processes in the Galactic field that couple different components of the foregrounds. This network is able to achieve significantly higher correlations across all scales and $f_{\rm{resid}}\simeq 0.376$. 
    \item We also trained the UNet on various input combinations with $T,E,B$ components to explore which provides the most information for $B$-mode reconstruction. We find that using just E-modes results in more accurate and consistent reconstructions than the small-scale B-modes alone. This makes sense given that E-modes probe fluctuations correlated with the GMF~\citep{Clark_2021} and contains information at all angular scales. Meanwhile, temperature, which can be understood to only modulate the overall intensity of the perturbations, provides the least amount of information compared to $T$ and $B_S$. 
\end{enumerate}

Next, we simultaneously leverage multi-frequency and inter-scale correlations by training a hybrid network. The inputs consist of ILC foreground estimates $\hat{F}^{\rm{ILC}}_i$ at 4 frequencies, small-scale B-modes $B_S$ at 4 frequencies, and optionally T and E modes at 4 frequencies each. Hence, the maximal network involves a total of 16 input channels. The targets are ILC residuals $\Delta F^{\rm{ILC}}$, which are then subtracted from the ILC reconstruction to get the ultimate, UNet-improved CMB reconstruction. This advantageously combines frequency-dependent SED correlations, scale-dependent structural correlations, and cross-correlations between different CMB observables ($T$, $E$, and $B$) in one network (see Table~\ref{tab:multifreq_performance}):
\begin{enumerate}
    \item The 4-channel spatial ILC achieves a baseline of $f_{\rm{resid}} = 2.69\times10^{-3}$. This is reduced by a network trained on ILC foreground estimates $\hat{F}^{\rm{ILC}}_i$ to $f_{\rm{resid}} = 1.00\times10^{-3}$ ($\sim 2.7\times$ lower), removing residual anisotropic power that ILC could not mitigate. This agrees with the results of~\citep{mccarthy24_ml}.
    \item {Adding small-scale B-modes ($\hat{F}^{\rm{ILC}}_i + B_S$) further improves cleaning to $f_{\rm{resid}} = 4.71\times10^{-4}$, showing that inter-scale information captures anisotropic structure beyond frequency correlations alone.}
    \item {The maximal 16-channel network ($\hat{F}^{\rm{ILC}}_i + B_S + T + E$) reaches $f_{\rm{resid}} = 3.62\times10^{-4}$, leaving only $\sim0.036\%$ of the large-scale foreground power and demonstrating that $T$ and $E$ provide a modest but non-redundant further gain.}
\end{enumerate}

Going forward, several important limitations must be addressed before these methods can be applied to observational data. First, our networks are trained and evaluated on one physical model of the Galactic sky, via the \textsc{DustFilaments} simulations. Hence, a pressing follow-up is to investigate robustness to other foreground models, such in the \textsc{PySM3} suite  \citep{thorne2017pysm} or thermal emission maps from the {\textsc{Vansyngel} simulations~\citep{Vansyngel_2018}}, with different assumptions about dust morphology and Galactic magnetic field structure. This will help determine whether the networks are learning physical foreground characteristics or simply overfitting to the simulation. To achieve robust results across different models, {transfer learning~\citep[e.g.,][Sec.~15.2]{Goodfellow-et-al-2016} and domain adaptation techniques~\citep[e.g.,][]{Long_2013_ICCV}} could also help bridge the gap between simulated and real data, while systematic studies of which simulations are most realistic or representative could guide training set choice. Finally, to obtain constraints on cosmology, it will be necessary to train on simulations with varying levels of primordial non-Gaussianity and different tensor-to-scalar ratio values ($r$). 

Other considerations on systematics can be made to improve the realism of the pipeline for observations. This includes training on larger sky areas to better capture large-scale correlations that may be missed in patch-based approaches. It is also critical to account for instrumental effects, including realistic noise models, beam convolution, and detector configurations. 

Moreover, there are several other physical phenomena that can affect foreground removal. For instance, frequency decorrelation, generated by flux ratios that vary across the sky and lines of sight~\citep[e.g.,][]{2021A&A...647A..16P}, can cause maps at different frequencies to not be perfectly correlated. The \textsc{DustFilaments} simulations model this by generating a random modified blackbody SED for each
filament~\citep{Herv_as_Caimapo_2022}. It would be interesting to explore how this effect impedes the multi-frequency networks' performance at varying degrees of decorrelation. Additionally, incorporating de-lensing templates will be a useful step in the pipeline, as gravitational lensing converts E-modes to B-modes and must be properly accounted for. Although our simulations omit gravitational lensing of the CMB, its inclusion need not break the networks' signal-preserving property. Weak lensing converts $E$-modes into $B$-modes through non-Gaussian mode coupling~\citep{LEWIS_2006}, and the resulting lensing $B$-modes can be treated as an additional secondary contaminant on top of the primordial signal. {Moreover, on the angular scales and frequencies relevant here, the power of lensing-induced $B$-modes is negligible compared to that of polarized Galactic dust}~\citep{planck2016polstats, planck_dust}.

It would also be fruitful to explore alternative strategies for incorporating ML reconstructions into ILC frameworks, such as iterative refinement approaches that alternate between ILC and ML corrections, and compare performance with other variants of ILC methods. {We present one way to do this in Appendix~\ref{sec:results_ilc_enhanced} where we use the UNet foreground prediction as an additional ILC channel.} The methods discussed in this work could also be extended to other foreground components beyond thermal dust, such as galactic synchrotron emission or extragalactic sources, further expanding their applicability to next-generation CMB experiments. 

Ultimately, standard foreground removal techniques that rely solely on frequency correlations leave a surplus of untapped spatial and cross-component information. The ML methods discussed here provide a promising avenue for anisotropic foreground mitigation that takes advantage of the additional information available across scale in temperature and polarization maps, while still ensuring that the signal of interest is preserved.  The complementary nature of single-frequency, inter-scale methods versus multi-frequency approaches offers flexibility for experiments with different observational constraints. As future surveys strive for tighter constraints on inflationary parameters, combining signal-preserving ML with multi-frequency algorithms offers a reliable and promising way towards optimal removal of complex CMB foregrounds and statistically improve the detection of primordial B-modes.

\appendix
\section{Network Architecture and Training Procedure}\label{app:nn_archi}

\subsection{Architecture}\label{app:arch_spec}
The UNet architecture follows the encoder-decoder structure of \citep{ronneberger2015unet}, consisting of 6 encoder levels and decoder levels, with feature channel dimensions of $[32, 64, 128, 256, 512, 1024]$. The encoder progressively down-samples the input from $128 \times 128$ to $4 \times 4$ pixels (bottleneck), while the decoder up-samples back to $128 \times 128$ pixels using transpose convolutions (\textsc{TransConv}) and skip connections. 

\begin{table}[ht!]
\centering
\caption{UNet architecture structure showing encoder and decoder levels with spatial dimensions, feature channels, and operations. The number of input channels, $N_{\rm{in}}$, depends on the number of frequencies and CMB components ($T,E,B$) used. The encoder progressively down-samples from $128 \times 128$ to $4 \times 4$ pixels, while the decoder up-samples back to $128 \times 128$ pixels. Each level shows the spatial size and channels after the \textsc{DoubleConv} operation. Skip connections between the corresponding layers of the encoder and decoder concatenate the downsampled and upsampled features channel-wise before each decoder \textsc{DoubleConv} block.}
\label{tab:unet_architecture}
\small
\begin{tabular}{clccc}
\toprule
\textbf{Level} & \textbf{Path} & \textbf{Spatial Size} & \textbf{Channels} & \textbf{Operation} \\
\midrule
0 & Input & $128 \times 128$ & $N_{\text{in}}$ & --- \\
\midrule
1 & Encoder & $128 \times 128$ & 32 & DoubleConv \\
 & & $64 \times 64$ & 32 & AvgPool($2 \times 2$) \\
\midrule
2 & Encoder & $64 \times 64$ & 64 & DoubleConv \\
 & & $32 \times 32$ & 64 & AvgPool($2 \times 2$) \\
\midrule
3 & Encoder & $32 \times 32$ & 128 & DoubleConv \\
 & & $16 \times 16$ & 128 & AvgPool($2 \times 2$) \\
\midrule
4 & Encoder & $16 \times 16$ & 256 & DoubleConv \\
 & & $8 \times 8$ & 256 & AvgPool($2 \times 2$) \\
\midrule
5 & Encoder & $8 \times 8$ & 512 & DoubleConv \\
 & & $4 \times 4$ & 512 & AvgPool($2 \times 2$) \\
\midrule
6 & Bottleneck & $4 \times 4$ & 1024 & DoubleConv \\
 & & $4 \times 4$ & 2048 & DoubleConv \\
\midrule
5' & Decoder & $8 \times 8$ & 512 & TransConv($2 \times 2$) + Skip + DoubleConv \\
\midrule
4' & Decoder & $16 \times 16$ & 256 & TransConv($2 \times 2$) + Skip + DoubleConv \\
\midrule
3' & Decoder & $32 \times 32$ & 128 & TransConv($2 \times 2$) + Skip + DoubleConv \\
\midrule
2' & Decoder & $64 \times 64$ & 64 & TransConv($2 \times 2$) + Skip + DoubleConv \\
\midrule
1' & Decoder & $128 \times 128$ & 32 & TransConv($2 \times 2$) + Skip + DoubleConv \\
\midrule
Output & Final & $128 \times 128$ & 1 & Conv($1 \times 1$) \\
\bottomrule
\end{tabular}
\end{table}

Each encoder and decoder block contains a \texttt{DoubleConv} module consisting of two sequential $3 \times 3$ convolutions that maintain spatial dimensions, followed by group normalization after each convolution~\citep{wu2018group}, and LeakyReLU activation layer with negative slope $\alpha = 0.01$. Encoder downsampling uses average pooling between encoder layers, while the decoder upsampling block uses transpose convolution between its layers. Encoder feature maps are concatenated channel-wise with upsampled decoder features before each decoder \texttt{DoubleConv} block via skip connections, with bilinear interpolation handling size mismatches when concatenating skip connections (for odd spatial dimensions). 

For more stable training and faster convergence, we adapt target-mean bias initialization where we manually initialize the output layer bias to the pixel-wise mean of the target patches in the training set to provide a sensible starting point (see next section for alternative normalization). Meanwhile, the CNN layers use Kaiming normal initialization~\citep{he2015delving}. Table~\ref{tab:unet_architecture} summarizes the UNet architecture, showing encoder and decoder layers with their corresponding spatial dimensions, feature channels, and operations.

\subsection{Training Procedure}\label{app:training_config}

During training, we use the \textsc{Adam} optimizer \citep{kingma2017} with initial learning rate of $10^{-3}$ and weight decay of $5 \times 10^{-4}$. We also employ a \texttt{ReduceLROnPlateau} learning rate scheduler that reduces the learning rate by factor 0.5 when validation loss does not change over an interval of 10 epochs. We use batch sizes of 64 and train for a maximum 200 epochs, with early stopping if validation loss does not improve for 20 epochs. We then save the best model at the lowest validation loss for evaluation on test set patches.

Finally, we employ data augmentation to prevent overfitting and increase variability. This includes spatial transformations: rotations ($90^\circ$, $180^\circ$, or $270^\circ$) and flips (horizontal and/or vertical axis). We also tried normalizing the data using per-channel standardization with the training set mean and standard deviation to stabilize the dynamic range of the network inputs. We find that this gives similar performance compared to initializing the output layer bias as the training set mean, as mentioned in Appendix~\ref{app:arch_spec}, so we resort to using the latter for the reported results.

\section{DustFilaments Simulation Normalization}\label{app:simulation_normalization}

The \textsc{DustFilaments} simulations model galactic dust distribution by populating the foreground sky with GMF-aligned dust filaments. The model's three main components are: (1) \texttt{get\_MagField}, which generates a 3D magnetic field cube with an optional large-scale Galactic component and an isotropic random component; (2) \texttt{get\_FilPop}, which creates a random population of filaments aligned with the magnetic field; and (3) \texttt{Paint\_Filament}, which paints each filament into HEALPix maps of temperature and polarization (T, Q, U Stokes parameters) at specified frequencies. The filament population naturally produces correlated structures on all scales. At scales ($\ell \lesssim 50$) for $Q, U$ components, where the model uses a \textit{Planck}-based template filled in from observations, the frequency extrapolation follows a standard modified blackbody law with spatially-varying the parameters  $\beta_{\text{dust}}$ (the dust spectral index), and $T_{\text{dust}}$ (the effective dust temperature),  from Planck GNILC. Note that a template is not used for the temperature map, which is generated solely from filaments all the way to the largest scales.

The fiducial simulations are normalized to match \Planck 353\,GHz observations following the calibration procedure described in \citep{Herv_as_Caimapo_2022}. The normalization occurs in two stages: first, the T,Q,U maps are calibrated to physical units ($\mu$K) by matching the small-scale polarization power spectra (EE and BB) to the \Planck 353\,GHz dust spectra from \citep{planck2018componentsep}. Specifically, a $\chi^2$ minimization is performed in the multipole range $280 < \ell < 600$, where the one-filament contribution dominates, using the Planck-measured power spectra in the LR71 mask with error bars from simulations. The filament density is then calibrated to match the large-scale temperature power spectrum, with the total number of filaments for the full-sky chosen to reproduce the \textit{Planck} $\mathcal{D}_\ell^{\text{TT}}$ spectrum in the range $2 \leq \ell < 600$. To generate maps at frequencies other than the 353\,GHz anchor band, the \textsc{DustFilaments} model uses a modified blackbody (MBB) spectral energy distribution to extrapolate to other frequencies. 

\section{Enhanced ILC with UNet Prediction as Additional Channel}\label{sec:results_ilc_enhanced}

\begin{figure}
    \centering
    \includegraphics[width=1\linewidth]{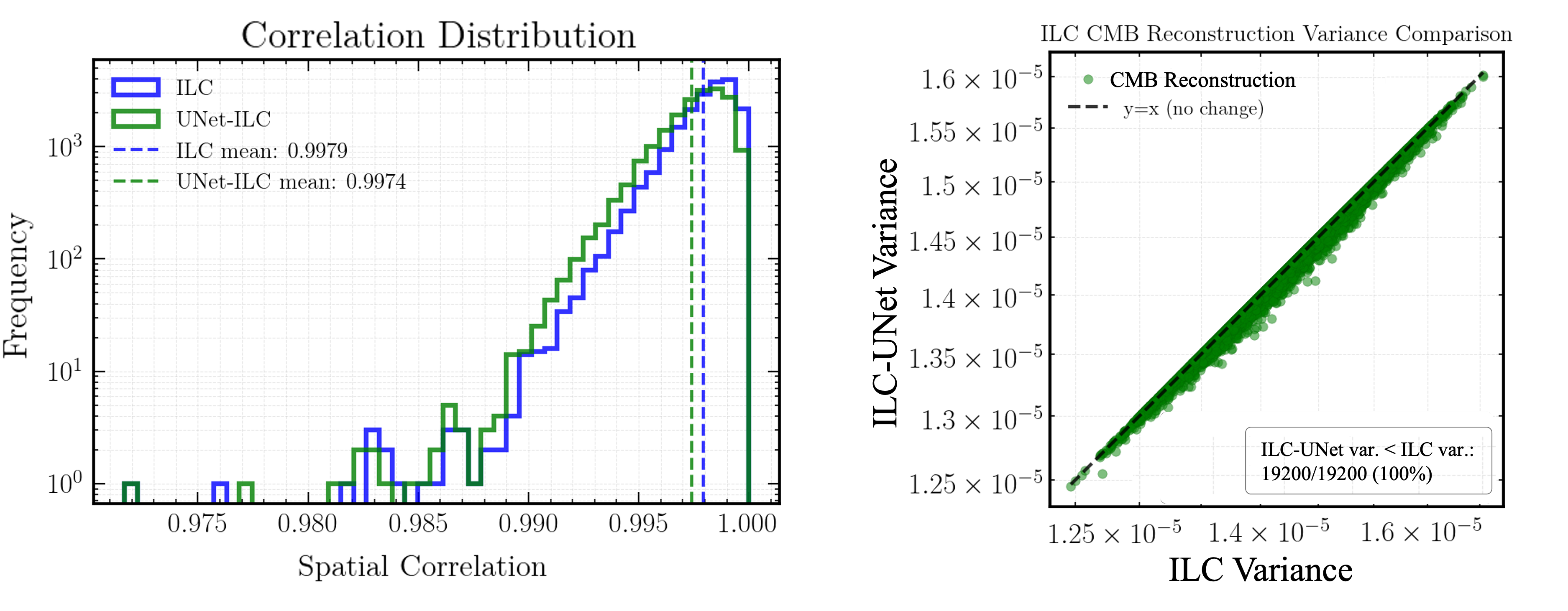}
    \caption{Results of the CMB reconstructions obtained using an ILC augmented with a synthetic channel derived from the UNet foreground prediction trained on $T+E+B_S$ inputs (see Section~\ref{sec:interscale_t_e}).
    \textbf{Right:} Comparison of the spatial correlation achieved by the ILC with and without the synthetic channel. The inclusion of the UNet-based channel does not lead to a significant improvement, and the mean spatial correlation is slightly reduced relative to the standard ILC solution. Similar behavior is observed for the harmonic correlation coefficient.
    \textbf{Left:} Comparison of the reconstruction variance for the ILC with and without the additional channel. As expected from the variance-minimization framework, the UNet-enhanced ILC solution achieves variance that is equal to or lower than that of the standard ILC.
    Overall, the absence of improvement in correlation indicates that the synthetic channel does not provide additional independent information for determining the ILC weights. Consistent results are obtained when using UNet predictions from the $B_S$-only network discussed in Section~\ref{sec:interscale_b}.}
    \label{fig:unet-ilc-var}
\end{figure}
While the UNet-based foreground reconstruction methods presented in Sections~\ref{sec:results_interscale} and~\ref{sec:results_combined} demonstrate significant improvements over baseline ILC, we were also interested in exploring how ML predictions could complement conventional ILC algorithms. This motivated our investigation of an enhanced-ILC approach that treats the UNet-predicted CMB reconstruction as an additional, synthetic frequency channel -- effectively allowing the ILC algorithm to optimally combine the original multi-frequency observations with the UNet estimate of the foregrounds. Hence, we construct an enhanced ILC solution using the extended set of maps at the following frequencies: $[B_{95}, B_{145}, B_{220}, B_{270}, \hat{B}_{\text{220}}]$. Here, $\hat{B}_\text{220} = S + \hat{F}_L$ is the UNet's reconstructed CMB map (containing both the primary CMB signal at all scales, and the predicted large-scale foreground at 220 GHz. The ILC weights are then computed on this extended 5-channel set via variance-minimization under the constraint that the weights sum to unity (Eqn.~\ref{eq:ilc_weights}).

The motivation behind this can be better understood as follows. While traditional ILC optimizes the weights to build a linear combination of the given channels, it is unable to explicitly isolate or upweight non-Gaussianities and/or anisotropies in the maps. The synthetic channel therefore provides a complementary encoding of the foreground information such as spatially coherent anisotropic patterns, large-scale filamentary dust structures, and other processes in the interstellar medium. In the presence of noise, this his morphological dust \textit{template} thereby emphasizes large-scale coherent patterns that may be more clearly separated from noise and better represented than in the raw $B_{220}$ channel, even though both are derived from the same underlying observations and the latter cannot contain more information than the observed $B_{220}$ map. The UNet's non-linear modeling can also suppress noise components while preserving physically meaningful foreground structures. The worst-case scenario is that the UNet reconstruction only contains information that is redundant with the linear combination of existing channels, resulting in no change to the standard ILC solution.

We evaluate the effectiveness of the enhanced ILC by comparing its reconstruction correlations with those of the conventional ILC. This is shown in the left plot of Fig.~\ref{fig:unet-ilc-var} where the histograms reveal that the UNet-enhanced ILC does not yield improved reconstructions, -- achieving a similar distribution of $\rho$ across test patches. We find similar results for harmonic $\rho(\ell)$. Nevertheless, the UNet-enhanced ILC always achieves a lower or equal variance compared to the traditional ILC, as shown in the right panel. This is expected since the ILC weights minimize variance by construction and cannot increase it with the inclusion of additional channels.

\begin{figure}[ht!]
    \centering
    \includegraphics[width=0.65\linewidth]{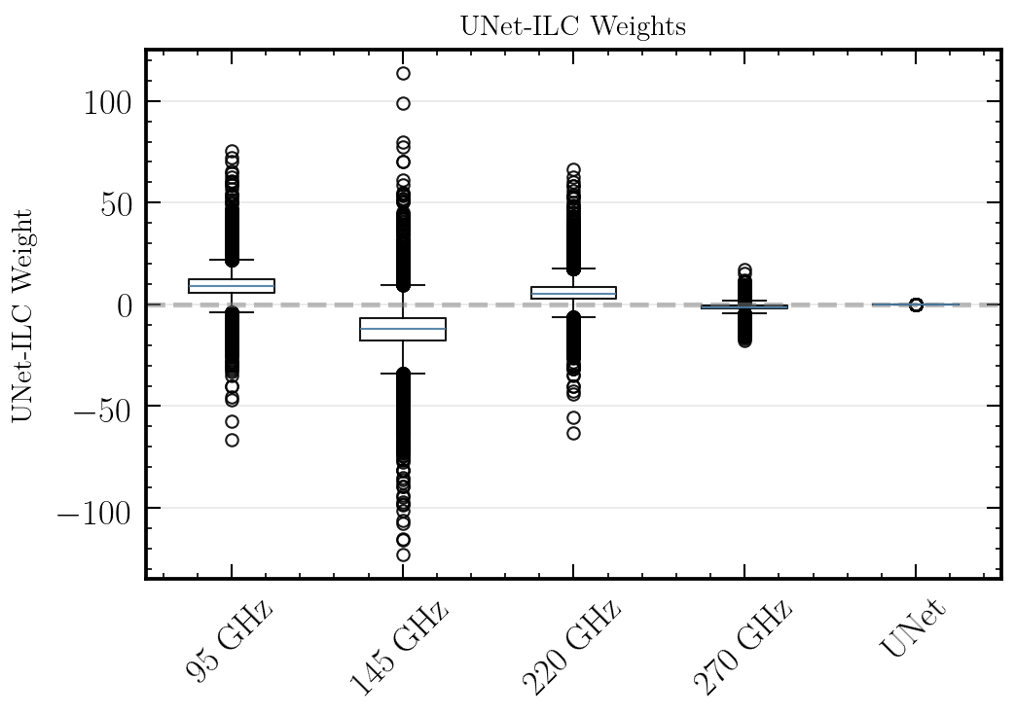}
    \caption{Distribution of ILC weights obtained when including the UNet's CMB reconstruction as an additional channel in the ILC optimization. The UNet estimate receives negligible weight in the final solution, indicating that it contributes little independent information beyond that already present in the existing channels.}
    \label{fig:unet-ilc-weights}
\end{figure}

To better understand how the enhanced ILC utilizes the available inputs, we examine the distribution of the ILC weights assigned to each channel. We find that the additional synthetic channel is assigned a negligible weight, close to zero, as shown in Fig.~\ref{fig:unet-ilc-weights}. This indicates that the synthetic channel does not provide independent information that is useful for determining the ILC weights. At the same time, the inclusion of this channel slightly modifies the weights assigned to the original channels, which may account for the modest degradation in correlation observed for the UNet-enhanced ILC solution. We leave a more detailed investigation of alternative strategies for incorporating the UNet foreground estimate into ILC-based methods to future work. For instance, the strong ILC performance may be saturated by the spatial correlations across frequency, which can be weakened in the presence of realistic noise or frequency decorrelation. Using more synthetic channels at other frequencies may also help.

\section*{Acknowledgments}
This work is not an official Simons Observatory paper. CHC acknowledges ANID for grant BASAL CATA FB210003. Miles Cranmer is grateful for support from the Schmidt Sciences AI2050 Early Career Fellowship and the Isaac Newton Trust. HS acknowledges that this material is based upon work supported by the National Science Foundation Graduate Research Fellowship Program under Grant No. DGE 2140743. Any opinions, findings, and conclusions or recommendations expressed in this material are those of the author(s) and do not necessarily reflect the views of the National Science Foundation.

\bibliographystyle{apsrev4-2}
\bibliography{references}

\end{document}